\documentclass[10pt, conference, letterpaper]{IEEEtran}
\usepackage{geometry}
\usepackage{algorithm}
\usepackage{cite}
\usepackage{graphicx}
\usepackage{amsmath}
\usepackage{amsthm}
\usepackage{mathtools}
\usepackage{algorithmic}
\usepackage{array}
\usepackage{subcaption}
\usepackage{gensymb}
\usepackage{amssymb}
\usepackage{multicol}
\usepackage{multirow}
\usepackage{float}
\usepackage{framed}
\usepackage{xcolor}
\usepackage{pgfplots}
\pgfplotsset{compat=newest}
\usetikzlibrary{plotmarks}
\usetikzlibrary{arrows.meta}
\usepgfplotslibrary{patchplots}
\usepackage{grffile}
\newtheorem{theorem}{Theorem}[section]

\newcommand\numberthis{\addtocounter{equation}{1}\tag{\theequation}}

\newlength\tikzfigH
\newlength\tikzfigW

\usepackage{pgfplots}

\makeatletter
\def\blfootnote{\gdef\@thefnmark{}\@footnotetext}
\makeatother

\IEEEoverridecommandlockouts

\begin{document}
\title{ROBin:  Known-Plaintext Attack Resistant Orthogonal Blinding via  Channel Randomization }
\author{\IEEEauthorblockN{Yanjun Pan\IEEEauthorrefmark{1} \quad
Yao Zheng\IEEEauthorrefmark{2} \quad
Ming Li\IEEEauthorrefmark{1}}
\IEEEauthorblockA{\IEEEauthorrefmark{1}The University of Arizona, Tucson, AZ\\
\IEEEauthorrefmark{2}University of Hawai`i at M\=anoa, Honolulu, HI\\
Email: \IEEEauthorrefmark{1}\{yanjunpan,lim\}@email.arizona.edu \quad \IEEEauthorrefmark{2}yao.zheng@hawaii.edu }
\thanks{This work is partly supported by NSF grants CNS-1619728, CNS-1564477, ONR grant N00014-16-1-2650, ARO grant W911NF-19-1-0050.}}

\maketitle
\begin{abstract}
Orthogonal blinding based schemes for wireless physical layer security aim to achieve secure communication by injecting noise into channels orthogonal to the main channel and corrupting the eavesdropper's signal reception. These methods, albeit practical, have been proven vulnerable against multi-antenna eavesdroppers who can filter the message from the noise. The vulnerability is rooted in the fact that the main channel state remains static in spite of the noise injection, which allows an eavesdropper to estimate it promptly via known symbols and filter out the noise. Our proposed scheme leverages a reconfigurable antenna for Alice to rapidly change the channel state during transmission and a compressive sensing based algorithm for her to predict and cancel the changing effects for Bob. As a result, the communication between Alice and Bob remains clear, whereas randomized channel state prevents Eve from launching the known-plaintext attack. We formally analyze the security of the scheme against both single and multi-antenna eavesdroppers and identify its unique anti-eavesdropping properties due to the artificially created fast-changing channel. We conduct extensive simulations and real-world experiments to evaluate its performance. Empirical results show that our scheme can suppress Eve's attack success rate to the level of random guessing, even if she knows all the symbols transmitted through other antenna modes.
\end{abstract}

\IEEEpeerreviewmaketitle

%-------------------- Intro --------------------
\section{Introduction}
The ever-expanding wireless technology is pushing the limit of the network security infrastructure. Many wireless devices need to secure the communication channels between each other without pre-shared security context. Orthogonal blinding based physical-layer security \cite{negi2005secret,goel2008guaranteeing,liao2010qos,li2011safe,anand2012strobe,argyraki2013creating} has been widely considered as a promising candidate to provide confidentiality during wireless transmission without \textit{a priori} key exchange. Instead of relying on pre-shared secrets, orthogonal blinding achieves secure communications by transmitting artificial noise into the null-space of the receiver's channel and  corrupting the eavesdropper's reception. Its practicality supersedes other theoretical physical-layer methods, such as zero-forcing beamforming,  which relies on knowledge about the eavesdropper's channel. Security analysis proves that it can asymptotically approach the secrecy rate of zero-force beamforming against single-antenna eavesdroppers. However, further studies show that orthogonal blinding is not effective against a multi-antenna eavesdropper, who has sufficient spatial dimensions to separate the message from the artificial noise. Schulz and Zheng et al. \cite{schulz2014practical, ZhengHighlyEfficientKnownPlaintext2015, zheng2016profiling} demonstrated that an eavesdropper may leverage the known or low entropy symbols in the transmission to quickly train a decoding filter and recovers the rest of the transmission, an attack equivalent to the known-plaintext attack in cryptanalysis.

The root of this vulnerability is due to the fact that the artificial noise only changes the quality of the receiving signal but not the state of the channel. Specifically, the noise injected by the transmitter (Alice) can lower the signal-to-noise ratio (SNR) of the eavesdropper's (Eve's) channel. But it cannot change the channel states between she and Eve or she and the legitimate receiver (Bob). This limitation opens up a window for the known-plaintext attack. Assuming the channel state remains ergodic with its coherent time. Due to the increasingly sophisticated digital modulation methods, Alice can transmit a sequence of tens or hundreds of symbols within such a short period. Although these symbols are buried deep under the artificial noise, a fraction of known symbols among them would allow Eve with multiple antennas to compute the channel state information (CSI), using a common MIMO technique known as least square (LS) channel estimation, which is robust against channel noise. Once Eve estimated the CSI, she may use it to equalize the channel and remove the artificial noise during the rest of the coherent period.

Follow this line of reasoning, there are two ways to defend against the known-plaintext attack, assuming Alice cannot avoid transmitting known symbols. She can limit the number of symbols to transmit within each coherent time period, which limits the communication throughput. Or she can reduce the coherent time to thwart the known-plaintext attack. However, the coherent time is an intrinsic condition that depends on the channel multipath and Doppler spread, both of which are not subject to the manipulation of transmitting content. Therefore, it would appear there are no cogent methods to defend against the known-plaintext attack.

However, in this paper, we challenge this no-win scenario and propose an orthogonal blinding based physical-layer security method immune to the known-plaintext attack: Channel-\textbf{R}andomized \textbf{O}rthogonal \textbf{B}l\textbf{in}ding (ROBin). ROBin leverages a pattern reconfigurable antenna to vary the channel state at a per symbol or per frame rate, resulting in an artificially created fast-changing wireless channel unsuitable for the known-plaintext attack, for which can be viewed as one of the proactive/dynamic defense (or moving target defense) mechanisms. To prevent the antenna reconfiguration from affecting Bob, we design a compressive sensing based algorithm for Alice to estimate the angle-of-departure (AoD) distribution of the multipath environment and predict the CSI for a given reconfigurable antenna pattern. Based on the predicted CSI, Alice can equalize the channel for Bob via digital pre-coding before transmitting. As a result, the main channel state appears stable  to Bob but randomly changing from Eve's perspective.

We formally analyze the security of ROBin, by comparing the mutual information between Alice's transmission and Eve's reception, assuming the channel state has the Markov property and Eve knows the symbols transmitted via historical antenna modes but not the current one. The analysis shows that Eve gains little advantage from knowing previous symbols (as channel randomization reduces the channel correlation  and makes  the current channel state more unpredictable). We implement the key components of ROBin; validate    our theoretical analysis with extensive simulation and real-world experiments. Empirical results show that our scheme can suppress Eve's attack success rate to the level of random guessing, even if she knows all the symbols transmitted through other modes.

%-------------------- Relatedwork --------------------
\section{Related Work}
Physical-layer security was pioneered by Wyner's work on the wiretap channel \cite{wyner1975wire}, which leverages the \textit{channel advantage} for legitimate receivers over degraded eavesdroppers to guarantee secure transmission over wireless channels. 
In \cite{wyner1975wire}, the rate of secret communications is characterized by \textit{secrecy capacity}, which is shown to be the difference in the capacity of the receiver and the eavesdropper. Following Wyner's work, numerous studies based on various channel models ranging from basic Gaussian channels to complex MIMO wiretap channels have been proposed later \cite{csiszar1978broadcast,leung1978gaussian,parada2005secrecy,li2007secret,gopala2008secrecy,KhistiSecureTransmissionMultiple2010,KhistiSecureTransmissionMultiple2010a}. In particular, Khisti et al. \cite{KhistiSecureTransmissionMultiple2010,KhistiSecureTransmissionMultiple2010a}
showed the secrecy capacity bounds in the large antenna limit with full channel state information (CSI) assumption. Their works reveal an important result   that the achievable secrecy capacity can be significantly affected by the number of antennas of the eavesdropper. 
% For instance, to block secret communication, Eve only needs three times as many antennas as transceivers have. 
However, since those theoretical works often make unrealistic assumptions such as channel advantage, full channel knowledge, or independent and identically channel distribution, they are rarely adopted to evaluate the secrecy of real-world schemes.

On the other hand, various practical physical-layer secret communication schemes  have been proposed.
One example is the friendly jamming approach.
Gollakota et al. prevented unauthorized commands from being transmitted to implantable medical devices (IMDs) in \cite{gollakota2011they}. They assume that the attacker equipped with MIMO is unable to separate the legitimate and jamming signal, due to the close proximity between the jammer and the data source. Similarly, Shen et al. \cite{shen2013ally} designed another jamming technique where jamming signals are controlled with secret keys, so that they are recoverable to authorized devices but unpredictably interfering to unauthorized ones. The jammer and the authorized device are very close to each other in both schemes, and this design is found as vulnerable by Tippenhauer et al. in \cite{tippenhauer2013limitations}. When an attacker tactfully places
her antenna array, the transmitted data signal can be recovered by exploiting the phase offsets between received signal components.
% Orthogonal blinding \cite{anand2012strobe} proposed by Anand et al. is another example of physical-layer security schemes. 
Artificial noise injection strategy \cite{negi2005secret,goel2008guaranteeing,liao2010qos,li2011safe} is another example  \cite{negi2005secret,goel2008guaranteeing,liao2010qos,li2011safe}, it has drawn significant attention by the security community since first proposed by Goel and Negi. However, it also relies on the unrealistic assumption that the statistics of the eavesdropper's channel are known to the transmitter. Argyraki et al.  in \cite{argyraki2013creating} proposed a cooperative jamming strategy for group secret agreement. By injecting  artificial noise through beamforming, a group of legitimate users are enable to create a shared secret, that the eavesdropper obtains very little information. However, this approach limits the number of antennas the eavesdropper possesses, which can be vulnerable to powerful eavesdroppers. On the other hand, Anand et al. proposed the orthogonal blinding scheme where no channel information about the eavesdropper is required \cite{anand2012strobe}.
To defend against a single-antenna eavesdropper, the transmitter injects artificial noise into channels orthogonal to the legitimate receiver's channel so that the original signal intended for the receiver cannot be recovered from the signal and noise mixture. However, when the eavesdropper has multiple antennas, by exploiting the known parts of the transmitted signal such as frame preambles, Schulz et al. \cite{schulz2014practical} successfully implemented a known-plaintext attack against orthogonal blinding. With normalized least mean square algorithms, an adaptive filter was trained to separate transmitted messages from artificial noise.

The root cause of the vulnerability in orthogonal blinding is that the channel is assumed to be stable during the whole transmission period, so that the attacker is able to gather enough plaintexts for filter training, and this flaw can be amended with channel randomization approach.
In the literature, the channel randomization approach has been used for key generation, message confidentiality, and integrity protection. Aono et al. \cite{aono2005wireless} proposed a key generation and agreement scheme that blocks the eavesdropper from generating the same key as transceivers by increasing the fluctuation of the wireless channel with a smart antenna. Hassanieh et al. \cite{hassanieh2015securing} presented a secret transmission scheme for RFIDs randomizing both modulation and channel by rotating several directional antennas at the transmitter. Different from this work, their scheme is only applicable to single-antenna transmitters and does not use pre-coding. To defend against active man-in-the-middle attacks, Hou et al. \cite{hou2015message} and Pan et al. \cite{pan2017message} randomized the wireless channel with a fan and a reconfigurable antenna respectively to prevent  online  signal cancellation.  All these works show that channel randomization approach can be a powerful tool to enhance physical-layer security. However, the studies are still preliminary and a comprehensive scheme that is MIMO-compatible and secure against multi-antenna attackers is lacking. 

%-------------------- Model --------------------
\section{System and Threat Models}
\begin{figure*}[!t]
    \vspace{-15pt}
    \centerline{
    \includegraphics[width=0.8\textwidth]{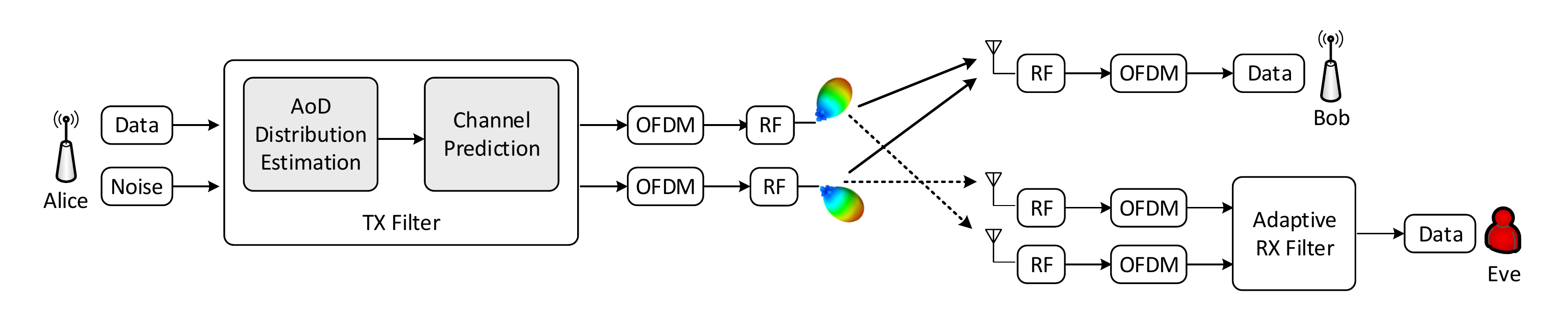}}
    \caption{Our system model illustrating the transmitter Alice, the legitimate receiver Bob and the passive eavesdropper Eve, where Alice is equipped with RA(s).}
    \label{fig:system}
    \vspace{-15pt}
\end{figure*}

Consider a MIMO-OFDM system shown in Fig. \ref{fig:system}, where the transmitter Alice aims at confidentially communicating with the receiver Bob through a wireless channel $\mathbf{H}_{AB}$, with the existence of a passive eavesdropper Eve. Denote the number of antennas for Alice, Bob and Eve as $n_{a}$, $n_{b}$ and $n_{e}$ respectively. 
The legitimate receiver Bob is equipped with regular omnidirectional antenna(s) (OAs), while the eavesdropper Eve can possess any types of antennas, including OAs, reconfigurable antenna(s) (RAs) and etc.. In particular, the transmitter Alice is equipped with RAs for channel randomization purpose, where an RA is an antenna capable of dynamically reconfiguring its antenna currents or radiating edges in a controlled and reversible manner \cite{bernhard2007reconfigurable}. Typically, an RA can swiftly reconfigurable its antenna profile including radiation pattern, polarization, frequency, and combinations of them. For example, Rodrigo et al. \cite{rodrigo2014frequency} presented an RA that has thousand of antenna modes and can be electronically switched within microseconds. From the receiver's perspective, the effect of the antenna profile is part of the CSI. Hence we can incorporate the impact of RA on the wireless channel into the channel model.

The wireless channel from Alice's $j$-th antenna to a receiver's $i$-th antenna ($(i,j)$-th receive-transmit pair) can be captured by a single complex number in the frequency domain, i.e. $h_{i,j}$, and the full CSI of transceivers can be represented by an array $\mathbf{H}$ with dimension  $n_b \times n_a$. Then the received signal $\mathbf{R}$ with dimension $n_b \times *$ can be expressed as:
\begin{equation}
    \mathbf{R} = \mathbf{H} \cdot \mathbf{D} + \mathbf{N}
\end{equation}
where $\mathbf{D}$ and $\mathbf{N}$ represents the transmitted data and the additive white Gaussian noise (AWGN), with dimension $n_a \times *$ and $n_b \times *$ respectively.
For the channel model, we consider a multipath channel. Recall that the effect of the antenna profile is also part of the CSI, to distinguish, we separate the CSI into the channel coefficient decided by the physical channel itself and the antenna part. Assuming that the channel is composed with $P$ multipaths, denote the physical channel coefficient part of $h_{i,j}$ as $h_{i,j}^{(phy)}$, then
\begin{equation}
    h_{i,j}^{(phy)}=\sum\limits_{l=1}^{P}L_{l}\alpha_{l}e^{-j\phi_{l}}
   \label{eq:h}
\end{equation}
where $L_{l}$ is the path loss of the $l$-th path, and $\alpha_{l}e^{-j\phi_{l}}$ is its fading parameter, here $\alpha_{l}$ and $\phi_{l}$ are the amplitude and phase of the fading respectively. Similar to existing works \cite{anand2012strobe,schulz2014practical,zheng2016profiling}, the physical channel coefficient $ h_{i,j}^{(phy)}$ in our model is fixed during the channel coherent time. Then the multipath channel can be expressed with the distribution of angle-of-departure (AoD). According to the multipath model, a single transmission from the antenna propagates along multiple paths before reaching the receiver. Each signal that travels at a particular AoD along with different paths experiences a different amount of attenuation and phase shifts. Then the physical channel coefficient part expressed as  \eqref{eq:h} can be further extended as the summation of the CSI over all the departure directions, and only the CSI that in the direction of multipaths is non-trivial. The distribution of CSI over all possible AoDs is defined as the AoD distribution. Then,
\begin{equation}
    h_{i,j}^{(phy)}=\sum\limits_{d=1}^{D}\mathbf{a}_{i,j}(\theta_d)
   \label{eq:h2}
\end{equation}
where the angular space is discretized into $D$ unique, equally spaced angles $\{\theta_1,\theta_2, \dots, \theta_D \}$, and $\mathbf{a}_{i,j}$ is the AoD distribution of $(i,j)$-th receive-transmit channel.

With RA, various antenna modes are associated with different radiation patterns. When the antenna gain under antenna mode $u$ and angle-of-departure $\theta_{l}$ is denoted as $\mathbf{G}(u,\theta_d)$, then the CSI $h_{i,j}$ under antenna mode $u$ can written as:
\begin{equation}
  h_{i,j}(u)=\sum\limits_{d=1}^{D}\mathbf{G}(u,\theta_{d})\mathbf{a}_{i,j}(\theta_d)
  \label{eq:hh}
\end{equation}
%and we illustrate this multipath channel model in terms of the AoD distribution with the effect of antennas in Fig. \ref{fig:channel}.

%
%\begin{figure}[t]
%    \vspace{-15pt}
%    \centering
%    \includegraphics[width=0.3\textwidth]{channel2}
%    \caption{Multipath channel model in terms of the AoD distribution, with the effect of antennas}
%    \label{fig:channel}
%    \vspace{-15pt}
%\end{figure}

% \begin{equation}
%   H_{i,j}=G(u,\theta_{i,j}) \cdot \sum\limits_{l=1}^{P}L_{l}\alpha_{i}e^{-j\phi_{i}}
%   \label{eq:hh}
% \end{equation}

Same as \cite{anand2012strobe,schulz2014practical,zheng2016profiling}, the channel from Alice to Bob ($\mathbf{H}_{AB}$) is measured at Bob's side and can be sent back to Alice through an out-of-band (OOB) channel or rely on implicit feedback, but $\mathbf{H}_{AB}$ is unknown to Eve. And Eve's can measure the channel from Alice to her ($\mathbf{H}_{AE}$), and it is unknown to neither Alice nor Bob. 

%-------------------- Physicallayersecurity --------------------
\section{Review of Orthogonal blinding}
Orthogonal blinding is designed to achieve secure communication by injecting noise into channels orthogonal to the main channel and corrupt the eavesdropper's signal reception. Though it is not the best possible achievable scheme for physical layer security,  as it does not require \textit{a priori} key exchange nor full channel knowledge, orthogonal blinding has been widely considered as a promising candidate to provide confidentiality during wireless transmission. However, it has been proven vulnerable against multi-antenna eavesdropper capable of discerning the message from the noise. In this section, we provide a brief review of orthogonal blinding scheme and the cause of its vulnerabilities.

\subsection{Transmitter-Side Precoding}
The core technique behind orthogonal blinding is known as transmitter-side precoding. To achieve secure transmission, Alice stirs both message and artificial noise (AN) via precoding. So Bob receives the pure message, and Eve receives the mixture of the noise and message. Zero-forcing and orthogonal blinding are two physical layer security schemes to achieved by transmitter-side precoding. 

In zero-forcing, Alice aims to transmit within the null-space of Eve's channel, which requires the full knowledge of Eve's channel. Such condition is not practical for a passive eavesdropper. While in orthogonal blinding, Alice needs only to know Bob's channel and transmits the AN in the null-space of Bob's channel to prevent eavesdroppers from extracting the data. Due to the orthogonality, Bob is not affected by the AN. But any receiver, whose channel is different from Bob's, receives a mixture of the message and AN. If the AN  in the mixture is strong, the receiver cannot recover the message.

The channels orthogonal to Bob's can be computed with the Gram-Schmidt algorithm as mentioned in \cite{anand2012strobe,schulz2014practical,zheng2016profiling}. First, Alice computes the projection matrix:
\begin{equation}
    \mathbf{H}_p = \mathbf{H}_{AB}^H(\mathbf{H}_{AB}\mathbf{H}_{AB}^H)^{-1}\mathbf{H}_{AB}
\end{equation}
and randomly generates a complex uniform matrix $\mathbf{H}'_{AN}$ with dimension $(n_{a}-n_{b}) \times n_{a}$. Then the difference between $\mathbf{H}'_{AN}$ and the projection of $\mathbf{H}'_{AN}$ is:
\begin{equation}
    \mathbf{H}^{''}_{AN} = \mathbf{H}'_{AN}-\mathbf{H}'_{AN} \cdot \mathbf{H}_p
\end{equation}
by normalizing this difference, we can obtain $\mathbf{H}_{AN}$:
\begin{equation}
    \mathbf{H}_{AN} = \frac{\mathbf{H}^{''}_{AN}}{\| \mathbf{H}^{''}_{AN} \|}
\end{equation}
where each row in $\mathbf{H}_{AN}$ is orthogonal to any other row in itself and to every row in $\mathbf{H}_{AB}$.

Next, Alice precodes the message ($\mathbf{D}_B$) and artificial noise ($\textbf{AN}$) with the pseudo-inverse of the matrix composed by $\mathbf{H}_{AB}$ and $\mathbf{H}_{AN}$, and obtain the transmitted signal $\mathbf{D}$ as:
\begin{gather}
    \mathbf{D} = \mathbf{F}_A\begin{pmatrix}
        \mathbf{D}_B\\
        \textbf{AN}
    \end{pmatrix} 
    \label{eq:data}
\end{gather}
where $\mathbf{F}_A$ is the transmit filter represented as:
\begin{gather}
    \mathbf{F}_A = \begin{pmatrix}
        \mathbf{H}_{AB}\\
        \mathbf{H}_{\text{AN}}
    \end{pmatrix}
    ^{H}
    \Bigg(
        \begin{pmatrix}
            \mathbf{H}_{AB}\\
            \mathbf{H}_{\text{AN}}
        \end{pmatrix}
        \begin{pmatrix}
            \mathbf{H}_{AB}\\
            \mathbf{H}_{\text{AN}}
        \end{pmatrix}
    ^{H}
    \Bigg)^{-1}
    \label{eq:F_A}
\end{gather}
Correspondingly, the received signal for Bob and Eve is:
\begin{equation}
    \begin{pmatrix}
    \mathbf{R}_B \\ \mathbf{R}_E
    \end{pmatrix} = 
    \begin{pmatrix}\mathbf{H}_{AB} \\ \mathbf{H}_{AE}
    \end{pmatrix} \cdot 
    \mathbf{D} + \mathbf{N}
\end{equation}
% where
% \begin{equation}
%     \mathbf{R}_E = \mathbf{H}_{AE} \cdot \mathbf{F}_A \cdot \begin{pmatrix} \mathbf{D}_B \\ \textbf{AN} \end{pmatrix} + \mathbf{N}
% \end{equation}

%-------------------- Attack --------------------
\subsection{Known-Plaintext Attack}
Anand et al. \cite{anand2012strobe} showed that single antenna eavesdroppers cannot recover the message with her reception, however, Schulz and Zheng et al. \cite{schulz2014practical,zheng2016profiling} showed that by exploiting the known parts or low entropy parts of the transmitted signal, the known-plaintext or ciphertext-only attack is possible in practice. Specifically, Schulz et al. introduced a practical known-plaintext attack for orthogonal blinding scheme. Unlike the typical assumption in the literature which assumes that the transmitted signal is fully unknown to the eavesdropper, Schulz argued that Eve can utilize the well-known protocols or addresses fields to guess part of the transmitted signal, so that some plaintext-ciphertext pairs are known to the eavesdropper, which is similar to the known-plaintext attack in cryptography. Then the eavesdropper can use the known plaintexts to train an adaptive filter for AN suppression. Ideally, the receive filter $\mathbf{F}_E$ is:
\begin{equation}
    \mathbf{F}_E = \mathbf{F}_A^{-1} \cdot \mathbf{H}_{AE}^{-1}
\end{equation}
In practice, Eve estimates $\mathbf{F}_E$ as $\hat{\mathbf{F}}_E$ with some known plaintexts $\mathbf{D}_B$ through iterative process. That is, Eve minimizes the mean square error between the estimated data and the known plaintexts:
\begin{equation}
    \min\limits_{\hat{\mathbf{F}}_E}~ E|\mathbf{D}_B - \hat{\mathbf{F}}_E \cdot \mathbf{R}_E|^2
\end{equation}
There are several iterative training algorithms for this problem, but in general, for a fixed transmit filter $\mathbf{F}_A$, multiple symbols are required to obtain a good adaptive filter at Eve's side due to the iterative training procedure. In \cite{schulz2014practical}, even with good training technique and parameter setting, $20 - 30$ training symbols are required when the ratio of transmitted AN to data is fairly low.

%-------------------- Scheme --------------------
\section{Robin: Channel-Randomized Orthogonal Blinding}
The vulnerability of preliminary orthogonal blinding results from the unchanged main channel, which allows the eavesdropper to estimate it via known symbols and filter the AN out. Actually, this flaw can be amended with the channel randomization approach, which is to actively randomize the wireless channel by introducing special antennas \cite{aono2005wireless,pan2017message}, antenna motions \cite{hassanieh2015securing} or artificial channel disturbance \cite{hou2015message}. Intuitively, when the wireless channel is rapidly randomized, 
% Eve can be blocked from gathering enough symbols for filter training, so that the message cannot be extracted from Eve's received signal. 
Eve can be blocked from gathering enough symbols for filter training. 
However, Eve can also explore the correlation between her channels and the main channel to estimate Bob's channel directly for message recovering. Results in \cite{popper2011investigation,pan2017message} showed that there is a strong correlation between two channels when the attacker is delicately positioned, and this correlation can be reduced with channel randomization \cite{pan2017message}. Hence, we propose a channel-randomized orthogonal blinding scheme which can be viewed as one of the proactive/dynamic defense (or moving target defense) mechanisms, to defend against known-plaintext attacks. 
We also show the benefits of reducing channel correlation to system security with the proposed metric in Sec. \ref{sec:theory}, which further supports our channel randomization approach.  

\subsection{Channel Prediction}
When the physical wireless channel remains unchanged, we randomize the wireless channel through rapid antenna mode switching, however, when the main channel changes, a new transmit filter is needed by Alice to guarantee the orthogonality between the message and noise subspaces. Traditionally, the main channel information $\mathbf{H}_{AB}$ is measured at Bob's side and sent back to Alice through an OOB channel or relying on implicit feedback.
% However, when the channel is randomized frequently, it is to costly for the transceivers to measure channels and get feedback every time, 
However, when the channel is randomized frequently, it becomes too costly, 
which makes the channel measurement a major challenge for orthogonal blinding based schemes. To solve this problem, we introduce a compressive sensing based channel prediction algorithm for Alice to cancel the channel changing effect to Bob. 

\subsubsection{AoD Estimation}
As the physical channel coefficient part is assumed as unchanged within the channel coherent time, it implies a stable AoD distribution correspondingly. To predict the CSI under different antenna modes, the distribution of AoD is estimated first to capture the physical wireless channel, and the effect of the antenna is added as in \eqref{eq:hh} for CSI prediction.

\subsubsection{Conventional AoD Estimation}
Traditionally, the distribution of AoD is estimated via MUSIC algorithm \cite{schmidt1986multiple}. To simplify, we describe it with a uniform linear array (ULA), with $M$ identical antenna elements arranged along a line with uniform spacing. Assume that there are $L$ multipath signals $S_1,S_2,\dots,S_L$ arriving. The matrix representation of the received signal at the array can be represented as:
\begin{equation}
    \mathbf{J} = \mathbf{A}\mathbf{S} + \mathbf{N} \label{eqJ}
\end{equation}
where $\mathbf{J}$ is the $M \times 1$ received signal, $\mathbf{S}$ is the $L \times 1$ signal source and $\mathbf{A}$ is the $M \times L$ steering vector matrix.

The basic idea of MUSIC algorithm is to implement eigenvalue decomposition of the received signal covariance matrix:
\begin{align}
    \mathbf{\Phi}_J & = E[\mathbf{J}\mathbf{J}^H] \label{eqRJ} \\
           & = \mathbf{A}\mathbf{\Phi}_S\mathbf{A}^H+\mathbf{\Phi}_N \label{eqRJ2}\\
           & = \mathbf{Q}_S\sum {\mathbf{Q}_S}^H+\mathbf{Q}_N\sum {\mathbf{Q}_N}^H
\end{align}
where $\mathbf{\Phi}_S$ and $\mathbf{\Phi}_N$ are the correlation matrix for the signal and noise respectively. Decomposing \eqref{eqRJ2} results in $M$ eigen values out of which the larger $L$ eigenvalues correspond to the multipath signals, where $\mathbf{Q}_S$ and $\mathbf{Q}_N$ are the basis of signal and noise subspaces respectively. Then by exploiting the orthogonality between the signal and noise subspaces, the direction of the arrived angles can be represented as:
\begin{equation}
    \theta_{MUSIC} = \text{argmin}~ \boldsymbol{\beta}^H(\theta)\mathbf{Q}_N{\mathbf{Q}_N}^H\boldsymbol{\beta}(\theta)
\end{equation}
% which is equivalent to obtain peaks in the spectral estimation:
% \begin{equation}
%     P_{MUSIC}=\frac{1}{\boldsymbol{\beta}^H(\theta)\mathbf{Q}_N{\mathbf{Q}_N}^H\boldsymbol{\beta}(\theta)}
% \end{equation}

However, as the MUSIC algorithm was mainly proposed for radio direction finding, the distribution obtained from it is only about the magnitude of CSI, which is not the AoD distribution we need. 
Hence, this algorithm is not applicable to our problem.
% Hence the traditional MUSIC algorithm is not applicable to our problem.

\subsubsection{Compressive AoD Estimation with RA}
Intuitively, the easiest way to estimate the AoD distribution for a given channel is to transmit with $D$ different antenna modes, so that we can solve \eqref{eq:hh} directly. However, it is not practical to estimate through this linear algebra approach due to large $D$ (e.g. in our case $D = 360$). Fortunately, by exploiting the sparsity of AoD distribution, the problem is solvable even with a small number of training modes. 

% The first step of channel prediction is the AoD distribution estimation, where Alice transmits pilots to Bob with a certain number of training antenna modes, and Bob sends the measured CSI back to Alice. With the known antenna profile and the corresponding CSI, Alice is able to obtain the AoD distribution by solving the linear algebra problem expressed as  \eqref{eq:hCS}. Since $a(\theta)$ is a vector of length $D$, it is only recoverable when we have $D$ different measurements of $H^{(u)}$. However, it is not practical to estimate AoD distribution through linear algebra approach due to large $D$ (e.g. in our case $D = 360$). Fortunately, by exploiting the sparsity of AoD distribution, the problem is solvable even with a small number of training modes.

Previous works \cite{ghassemzadeh2004measurement,czink2007cluster} have shown that for a typical multipath environment, there are only 3-5 distinct directions are dominant components. In other words, when we look into the AoD distribution, only a small number of them contribute significantly to the CSI. With this sparsity property, we can recover the AoD distribution from only a small number of measurements. Specifically, we use compressive sensing technique \cite{candes2008introduction} to estimate AoD distribution. 

Compressive sensing is a sampling algorithm that capable of recovering sparse signals with much fewer samples than traditional sampling approaches. One of the basic problems is to recover a signal $\mathbf{x}$ from a $M \times 1$ observation $\mathbf{y}$, with a given $M \times N$ sensing basis $\mathbf{\Phi}$, where $M < N$ and the signal $\mathbf{x}$ has a sparse representation with a $N \times N$ representation basis $\mathbf{\Psi}$ and $N \times 1$ weighting coefficients $\mathbf{s}$: $\mathbf{x} = \mathbf{\Psi s}$.  Mathematically speaking, the problem is to get $\mathbf{x}$/$\mathbf{s}$ from    
$\mathbf{y} = \mathbf{\Phi x} = \mathbf{\Phi \Psi s}$. The problem is solvable when the largest correlation between any two elements of $\mathbf{\Phi}$ and $\mathbf{\Psi}$ is small, which is refereed to as incoherence \cite{candes2008introduction}.

For our problem, since the AoD distribution $\mathbf{a(\cdot)}$ is sparse itself, our presentation basis degrades to an identity matrix, but we can still use the compressive sensing formulation to solve it. When training modes are randomly selected, the incoherence condition is roughly satisfied and the AoD distribution of $(i,j)$-th receive-transmit channel can be recovered from the following compressive sensing formulation: 
\begin{align}
    \begin{split}
        \mathbf{a}_{i,j} & = \text{argmin}~||\mathbf{a}_{i,j}(\theta)||_1\\
        \text{s.t.}~ h_{i,j}(u) & = \sum\limits_{d = 1}^{D} G(u,\theta_d)\mathbf{a}_{i,j}(\theta_d), \quad 1\leq u \leq U
    \end{split}
\label{eq:CS_formulation}
\end{align}
where $\| \cdot \|_1$ represents the L1 norm and $U \ll D$ are the total number of antenna modes needed for AoD distribution recovery. Note that, Xie et al. presented an estimation algorithm of AoA distribution based on compressive sensing in \cite{xie2015hekaton}. However, since they use antenna array, the CSI they use for estimation is the composite CSI instead of the one between each receive-transmit antenna pair. 
% Besides, similar to  MUSIC, their AoA distribution only computes the magnitude of the CSI.

\subsubsection{Channel Prediction}
Once the AoD distribution is estimated with the above compressive sensing formulation, the CSI $h_{i,j}$ under any given antenna mode can be predicted with  \eqref{eq:hh}. When the AoD distribution of every CSI element in the main channel is estimated, the whole matrix $\mathbf{H}_{AB}$ can be predicted correspondingly. Note that, the physical wireless channel is stable only within the channel coherent time, once the physical channel changes, a new round of AoD distribution estimation is required. In practice, since carrier frequency offset or accurate external clocks such as GPS clocks can eliminate the impact of frequency and phase offset, the channel coherent time can be long. Then the channel prediction is applicable, and it reduces the overhead for channel sounding comparing with the orthogonal blinding scheme.

\subsection{Secure Transmission Scheme}
In short, our RA based secure transmission scheme comprises two phases that we summarize hereunder, and for clarification, we denote the set of whole antenna modes, training modes, and transmitting modes as $\mathcal{S}$, $\mathcal{S}_1$, $\mathcal{S}_2$ respectively.

1. \textbf{Training phase}: 
 (a) Alice selects a certain number of antenna modes as training modes ($\mathcal{S}_1$). For each training mode $u \in \mathcal{S}_1$, several pilots are sent for each receive-transmit antenna pair $(i,j)$ with time-division multiplexing;
 
 (b) Bob measures the corresponding CSI $h_{i,j}(u)$ and shares it with Alice through OOB or implicit feedback;
 
  (c) Alice estimates the corresponding AoD distribution following \eqref{eq:CS_formulation} and gets $\mathbf{a}_{i,j}$.

2. \textbf{Secure transmission phase}:
(a) Alice randomly selects a set of antenna modes from the complement of $\mathcal{S}_1$ as transmitting modes ($\mathcal{S}_2 \subseteq \mathcal{S} \backslash \mathcal{S}_1$);

(b)  For each transmitting mode $v \in \mathcal{S}_2$, Alice predicts the corresponding channel matrix to Bob as $\hat{\mathbf{H}}_{AB}(v)$ following \eqref{eq:hh}, then the transmit filter $\mathbf{F}_A$ is computed based on the predicted $\hat{\mathbf{H}}_{AB}(v)$ following \eqref{eq:F_A};
        
(c) Alice transmits the message $\mathbf{D}_B$ and AN as in \eqref{eq:data}. For each packet, Alice uses a different mode randomly chosen from above, and Bob demodulates/decodes the received signal to get the messages from the packets directly.
 %   \end{enumerate}
%\end{enumerate}

Note that, the training phase needs to be executed once for every channel coherent time period (which is inversely proportional to the maximum Doppler spread of the physical channel). During the secure transmission phase, Alice does not need to include any pilots/preamble in the packets due to the transmit filter that cancels the channel effect to Bob. 

%-------------------- Theory --------------------
\section{Security Analysis}\label{sec:theory}
In this section, we formally analyze the security properties of ROBin. To model ROBin, we define the CSI of a wireless channel, $\mathbf{H}(\cdot)$ as a function of discrete-time $t$ and antenna mode $u$. Under this definition, the CSI in ROBin behaves as a function $\mathbf{H}\left(t,u(t)\right)$, where $u$ changes for each time step. We further assume that a sequence of $\mathbf{H}\left(t, u(t)\right)$s, forms a Markov chain \cite{tan2000first}, such that $\mathbf{H}\left(T, u(T)\right)$ is independent of past CSIs, $\left\{\mathbf{H}\left(t, u(t)\right)\mid t < T-1 \right\}$, given $\mathbf{H}\left(T{-}1, u(T{-}1)\right)$. To quantify the security of ROBin, we derive the conditional mutual information between Eve's receiving signal at time $T$, $\mathbf{R}_E(T)$, and the pre-blinding message, $\mathbf{D}_B(T)$, assuming Eve knows all previous CSIs between Alice and Bob, $\left\{\mathbf{H}_{AB}\left(t,u(t)\right)\mid t = 0,...,T{-}1\right\}$, and all CSIs between Alice and herself, $\left\{\mathbf{H}_{AE}\left(t,u(t)\right)\mid t = 0,...,T\right\}$ (Sec. \ref{sec:theo1}). Finally, we verify the correctness of the proposed metric and explain the insights gained from the analytical results (Sec. \ref{sec:theo2}).

\subsection{Secrecy Leakage as Conditional Mutual Information}
\label{sec:theo1}
% Secrecy capacity is a commonly used and well-studied metric in theoretical physical layer security studies. Although the results shown in the literature are important, they are rarely used for practical physical layer scheme evaluations, since most of them rely on assumptions that are hard to satisfy in practice (e.g., all the channels are known by the transmitter or all  nodes, SNR and block length go to infinity, etc.). Alternatively, perfect secrecy is a notation often considered in cryptosystems, which is defined as the ciphertext reveals no information about the plaintext. Similarly, the definition of perfect secrecy is so strict that makes it hardly being used.

% Perfect secrecy is a well-known cryptosystem model. Typically, a cryptosystem is considered to have perfect secrecy if, for any message $X$ and any ciphertext $Y$, we have $p(X|Y) = p(X)$, which indicates that the attacker learns nothing from the ciphertext. However, the definition is so strict that makes the notation rarely used in practice. Alternatively, 
% Besides, the definitions of perfect secrecy and secrecy capacity are not capable of indicating the practical PHY-layer attack schemes proposed in recent years. For example, for the practical known-plaintext attack proposed in \cite{schulz2014practical}, the eavesdropper's knowledge about the legitimate channel is crucial to attack performance, while this part of the information is not included in above two metrics. 

To quantify eavesdropper's capacity under known-plaintext attacks in a way congruence with cryptanalysis, we consider the secrecy leakage as the conditional mutual information between the Eve's receiving signal and the pre-blinding message, given Eve has full knowledge of all previous CSIs via known symbols. That is, we assume that, as $t=T$, all the previously transmitted symbols, $\mathbf{D}(t),\ t = 0,...,T{-}1$, are known to Eve, which allows Eve to compute $\mathbf{H}_{AB}(t, u(t)),\ t = 0,...,T{-}1$. 

Let $\mathcal{H}(T)$ defines a set of previous CSIs up to time T:
\begin{equation}
    \mathcal{H}(T) = \left\{\mathbf{H}\left(t,u(t)\right) \mid t = 0,...,T \right\}.
\end{equation}

Assuming $\mathcal{H}_{AE}(T)$ and $\mathcal{H}_{AB}(T-1)$ are known to Eve. The secrecy leakage is defined as a conditional mutual information:
\begin{equation}
    I\left(\mathbf{D}_B(T);\mathbf{R}_E(T)\mid\mathcal{H}_{AB}(T-1),\mathcal{H}_{AE}(T)\right)
    \label{metric1}
\end{equation}

% In particular, to consider the worst case to legitimate pairs, we consider a much stronger practical known-plaintext attack than that in \cite{schulz2014practical}. We assume that at time $t$, all the historical data from $X_1$ to $X_{t-1}$ is known to the attacker, and the known-plaintexts under any transmit filter are enough for the attacker to infer the historical CSI of channel A-B from $h_{AB}(1)$ to $h_{AB}(u-1)$ precisely. With the knowledge of her own channel and Bob's historical channel, the leakage of the system is defined as:
% \begin{equation}
%     I(X_t;Y_t|h_{AB}(1),\cdots,h_{AB}(u-1),h_{AE}(1),\cdots,h_{AE}(u-1),h_{AE}(u))
%     \label{metric1}
% \end{equation}
For simplicity, we first consider a single antenna system, in which $\mathbf{H}\left(t,u(t)\right)$ reduces to a scalar function $\mathbf{h}\left(t,u(t)\right)$. and the pre-coding filter becomes the inverse of the main channel, e.g., $F_A(T) = h_{AB}^{-1}\left(T,u(T)\right)$. Note that all derivations below also apply to MIMO system, which we will discuss later. The received signal at Eve's side is:
\begin{equation*}
\begin{split}
    \mathbf{R}_E(T) & = h_{AE}\left(T,u(T)\right) \left(h_{AB}^{-1}\left(T,u(T)\right) \mathbf{D}_B(T)\right) + \mathbf{N} \\
    & \triangleq h_{AB}^{-1}\left(T,u(T)\right)  \mathbf{D}_B(T) + \mathbf{N},
    \label{eq:yt}
\end{split}
\end{equation*}
after Eve equalizes $h_{AE}\left(T,u(T)\right)$. Omitting  $\mathbf{N}$, Eq. \eqref{metric1} expands to 
\begin{equation*}
    I\left(\mathbf{D}_B(T);h_{AB}^{-1}\left(T,u(T)\right)  \mathbf{D}_B(T) \mid \mathcal{H}_{AB}(T-1),\mathcal{H}_{AE}(T)\right)
    \label{metric2}
\end{equation*}
To simplify the equation above, consider the conditional probability of $h_{AB}\left(T, u(T)\right)$ given  $\mathcal{H}_{AB}(T-1)$. Due to the Markov property,  
\begin{equation*}
\begin{split}
& \Pr\left[h_{AB}\left(T, u(T)\right)\mid \mathcal{H}_{AB}(T-1)\right] = \\
& \ \ \ \ \Pr\left[h_{AB}\left(T, u(T) \right) \mid h_{AB}\left(T-1, u(T-1)\right)\right].
\end{split}
\end{equation*}
As for the conditional probability of $h_{AB}\left(T, u(T)\right)$ given $\mathcal{H}_{AE}(T)$. Although $h_{AB}\left(t, u(t)\right)$ and $h_{AE}\left(t, u(t)\right)$ are mostly independent, they are correlated at the same time step, since the antenna pattern is the same for $h_{AB}(u)$ and $h_{AE}(u)$, resulting
\begin{equation*}
\begin{split}
& \Pr\left[h_{AB}\left(T, u(T)\right)\mid \mathcal{H}_{AE}(T)\right] = \\
& \ \ \ \ \Pr\left[h_{AB}\left(T, u(T) \right) \mid h_{AE}\left(T, u(T)\right)\right].
\end{split}
\end{equation*}
Based on these conditions, we have the following Theorem:
\begin{theorem}
Assuming the wireless channel has the Markov property, the secrecy leakage of ROBin can be simplified as\footnote{The proof of this Theorem is in Appendix}:
% footnote{The proof of this Theorem can be found in our technical report at https://tinyurl.com/y2t4njcx}
\begin{align}
    I\left(\mathbf{D}_B(T);\mathbf{R}_E(T) \mid \right. & \left. \mathcal{H}_{AB}(T-1),\mathcal{H}_{AE}(T)\right) =  \nonumber \\
    I\left(\mathbf{D}_B(T);\mathbf{R}_E(T) \mid \right. & \left. h_{AB}\left(T-1, u(T-1)\right), \right. \nonumber \\
    & \left. h_{AE}\left(T-1, u(T-1)\right), \right. \nonumber \\
    & \left. h_{AE}\left(T, u(T)\right)\right) = \nonumber \\
    I\left(\mathbf{D}_B(T);\mathbf{R}_E(T) \mid \right. & \left. \delta \mathcal{H}_{ABE}(T)\right),
\label{eq:secrecy_leakage_simplified}
\end{align}
where
\begin{align*}
\delta \mathcal{H}_{ABE}(T) = \left\{ \right. & \left. h_{AB}\left(T-1, u(T-1)\right), \right. \\
& \left. h_{AE}\left(T-1, u(T-1)\right), \right.\\
& \left. h_{AE}\left(T, u(T)\right)\ \right\}
\end{align*}
\label{corollary2}
\end{theorem}
\vspace{-15pt}
This simplification allows us to calculate the numerical secrecy leakage when all the possible values of discretize CSI are in a small range. Next we use numerical results to show the relationship between channel correlation and privacy leakage.

\subsection{Correctness and Insights}
\label{sec:theo2}
\subsubsection{Single-Antenna Eavesdropper}
\begin{figure}[t]
    \centering
    \includegraphics[width=0.25\textwidth]{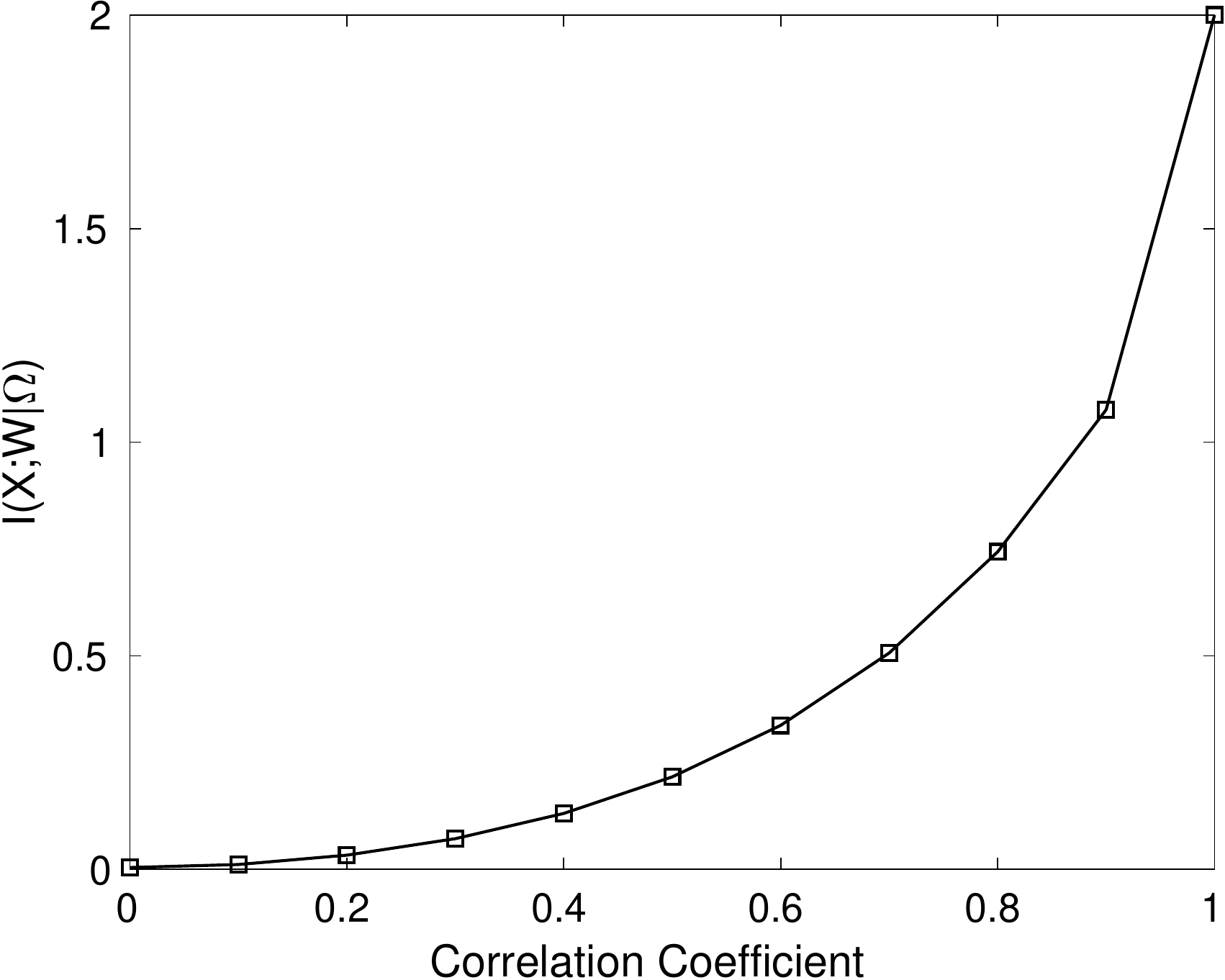}
    \caption{Secrecy leakage over the channel correlation coefficient between $\mathbf{H}_{AB}$ and $\mathbf{H}_{AB}$.}
    \label{fig:N}
    \vspace{-15pt}
\end{figure}

Alice can apply a reduced ROBin scheme without orthogonal blinding in a single-input and single-output (SISO) system, with Bob and Eve having one regular antenna and Alice having one reconfigurable antenna. To calculate the secrecy leakage, we first generate the CSI with the truncated Gaussian distribution in the range of $(-2,2)$, then we normalize its real (imaginary) part into four values, i.e. $\text{Re}[\delta \mathcal{H}_{ABE}(T)] \in \{\pm1.5,\pm0.5\}$. And for the message we consider 4QAM, namely that $\mathbf{D}_B(T) = x \in \{\pm1 + j, \pm 1 - j\}$, then the entropy of the message is $\mathrm{H}(\mathbf{D}_B(T)) = 2$. and $\text{Re}[\mathbf{R}_E(T)] \in \{\pm1.5,\pm0.5\}$, $|\mathbf{R}_E(T)| = 16$, $|(\mathbf{D}_B(T),\mathbf{R}_E(T),\delta \mathcal{H}_{ABE}(T))| = 256 \times 2^{10}$ correspondingly. Hence we set the number of the samples to 30 million, which is about 100 times of $|(\mathbf{D}_B(T),\mathbf{R}_E(T),\delta \mathcal{H}_{ABE}(T))|$. The calculated Eq. \ref{eq:secrecy_leakage_simplified} versus correlation coefficient between $\mathbf{H}_{AB}$ and $\mathbf{H}_{AB}$ is 
shown in Fig. \ref{fig:N}.

We can observe that the leakage increases with the increase of the correlation coefficient between $\mathbf{H}_{AB}$ and $\mathbf{H}_{AB}$, in other words, the information the eavesdropper gained decreases with the decrease of the correlation between $\mathbf{H}_{AB}$ and $\mathbf{H}_{AB}$. This result quantitatively verifies the motivation of our channel randomization strategy: the system becomes more secure after reducing the correlation between the two channels to the receiver and the eavesdropper. And results in \cite{pan2017message} shown that a reconfigurable antenna is capable of decreasing the correlation of two channels. Hence introducing a reconfigurable antenna to the system brings us two benefits: actively randomizing the wireless channel and reducing correlations among channels. 
% Besides, from the numerical results we can see that $I(X;\mathbf{W}|\Omega) \approx 0$ when correlation coefficient is 0, and $I(X;\mathbf{W}|\Omega) = H(X) = 2$ when correlation coefficient is 1, which are consistent with the lower bound and upper bound of this conditional mutual information.

%\subsubsection{MIMO System}
\subsubsection{Multi-Antenna Eavesdropper}
Alice can apply the full ROBin scheme with orthogonal blinding in a multi-input and single-output (MISO) or multi-input and multi-output (MIMO) system. Assume Eve has multiple antennas. 
For a given antenna mode, if the number of known symbols is less than the number of Alice's antennas $n_a$, Eve cannot find a unique decoding filter, because the least square problem for the LS channel estimation is underdetermined. However, the iterative decoding filter training process still provides Eve partial information about the message. And we use SER to evaluate this leakage in the simulation. When the number of known symbols is greater than $n_a$, the problem becomes overdetermined and allows Eve to identify the correct decoding filter. Nevertheless, as the number of known symbols do not accumulate when Alice reuses the same antenna mode at different channel coherent periods, as long as Alice switches the antenna mode faster than the duration of $n_a$ symbols, the secrecy leakage of our scheme is low regardless of the number of antennas Eve has.
\section{Performance Evaluation}
\begin{figure*}[t]
\begin{subfigure}[t]{0.23\textwidth}
\setlength{\tikzfigW}{0.9\linewidth}
\setlength{\tikzfigH}{0.9\linewidth}
% Minimal pgfplots version: 1.3
%  http://www.mathworks.com/matlabcentral/fileexchange/22022-matlab2tikz
\begin{tikzpicture}

\begin{axis}[%
width=\tikzfigW,
height=\tikzfigH,
at={(0.758333in,0.525in)},
scale only axis,
xmin=10,
xmax=40,
xtick={10, 20, 30, 40},
xlabel={No. of training modes},
ymode=log,
ymin=1e-08,
ymax=1,
ytick={ 1e-08,  1e-06, 0.0001,   0.01,    0.1},
yminorticks=true,
ylabel={SER},
label style={font=\small},
tick label style={font=\small},
legend style={at={(0.45,0.4)},anchor=south west,legend cell align=left,align=left,draw=white!15!black,font=\scriptsize}
]
\addplot [color=black,solid,line width=1.0pt]
  table[row sep=crcr]{%
10	0.0187450086805556\\
20	0.00150703125\\
30	3.90624999999978e-07\\
40	4.34027777777778e-08\\
};
\addlegendentry{$\text{Bob}_{\text{ROBin}}$};

\addplot [color=black,solid,line width=1.0pt,mark=star,mark options={solid}]
  table[row sep=crcr]{%
10	0.0603373263888889\\
20	0.0673694444444444\\
30	0.0612369357638889\\
40	0.0602316840277778\\
};
\addlegendentry{$\text{Eve}_{\text{ROBin}}$};

\addplot [color=black,dash dot,line width=1.0pt]
  table[row sep=crcr]{%
10	4.34027777777778e-08\\
20	4.34027777777778e-08\\
30	4.34027777777778e-08\\
40	4.34027777777778e-08\\
};
\addlegendentry{$\text{Bob}$};

\addplot [color=black,dashed,line width=1.0pt]
  table[row sep=crcr]{%
10	0.068654296875\\
20	0.0738588975694444\\
30	0.0780523003472222\\
40	0.0703971354166667\\
};
\addlegendentry{$\text{Eve}$};

\end{axis}
\end{tikzpicture}%
\end{subfigure}
\hspace{0.07\textwidth}
\begin{subfigure}[t]{0.23\textwidth}
\setlength{\tikzfigW}{0.9\linewidth}
\setlength{\tikzfigH}{0.9\linewidth}
% Minimal pgfplots version: 1.3
%  http://www.mathworks.com/matlabcentral/fileexchange/22022-matlab2tikz
\begin{tikzpicture}

\begin{axis}[%
width=\tikzfigW,
height=\tikzfigH,
at={(0.758333in,0.490972in)},
scale only axis,
xmin=1,
xmax=10,
xtick={ 1,  2,  4,  6,  8, 10},
xlabel={NDR},
ymode=log,
ymin=1e-08,
ymax=1,
ytick={ 1e-08,  1e-06, 0.0001,   0.01,      1},
yminorticks=true,
ylabel={Bob's SER},
label style={font=\small},
tick label style={font=\small},
legend style={at={(0.25,0.04)},anchor=south west,legend cell align=left,align=left,draw=white!15!black,font=\scriptsize}
]
% \addplot [color=black,dashed,line width=1.0pt]
%   table[row sep=crcr]{%
% 1	0.000986111111111111\\
% 2	0.00493433159722222\\
% 4	0.0260634548611111\\
% 6	0.0590388454861111\\
% 8	0.0982852864583333\\
% 10	0.137298567708333\\
% };
% \addlegendentry{$\text{SNR}_{\text{ROBin}}\text{=15dB}$}

% \addplot [color=black,dashed,line width=1.0pt,mark=x,mark options={solid}]
%   table[row sep=crcr]{%
% 1	0.000128602430555556\\
% 2	0.00164865451388889\\
% 4	0.00938823784722223\\
% 6	0.0204099392361111\\
% 8	0.0361276041666667\\
% 10	0.0567549479166667\\
% };
% \addlegendentry{$\text{SNR}_{\text{ROBin}}\text{=20dB}$}

\addplot [color=black,solid,line width=1.0pt]
  table[row sep=crcr]{%
1	0.000122743055555556\\
2	0.00112291666666667\\
4	0.00599483506944444\\
6	0.0107560329861111\\
8	0.0175454427083333\\
10	0.0252785590277778\\
};
\addlegendentry{$\text{SNR}_{\text{ROBin}}\text{=25dB}$}

\addplot [color=black,dashed,line width=1.0pt]
  table[row sep=crcr]{%
1	6.77083333333328e-06\\
2	0.00128888888888889\\
4	0.00580078125\\
6	0.00843862847222222\\
8	0.012955859375\\
10	0.0177450954861111\\
};
\addlegendentry{$\text{SNR}_{\text{ROBin}}\text{=30dB}$}

% \addplot [color=black,solid,line width=1.0pt]
%   table[row sep=crcr]{%
% 1	0.000699652777777778\\
% 2	0.00341688368055556\\
% 4	0.0201337239583333\\
% 6	0.0510474826388889\\
% 8	0.0879983506944444\\
% 10	0.125694965277778\\
% };
% \addlegendentry{$\text{SNR}_\text{OB}\text{=15dB}$};

% \addplot [color=black,solid,line width=1.0pt,mark=x,mark options={solid}]
%   table[row sep=crcr]{%
% 1	4.60069444444446e-05\\
% 2	0.000383159722222222\\
% 4	0.00349991319444444\\
% 6	0.01133359375\\
% 8	0.0246471788194444\\
% 10	0.0419066840277778\\
% };
% \addlegendentry{$\text{SNR}_\text{OB}\text{=20dB}$};

\addplot [color=black,dash dot,line width=1.0pt]
  table[row sep=crcr]{%
1	8.68055555555136e-08\\
2	7.20486111111118e-06\\
4	0.000420486111111111\\
6	0.00194926215277778\\
8	0.00467591145833333\\
10	0.00904301215277778\\
};
\addlegendentry{$\text{SNR}_\text{OB}\text{=25dB}$};

\addplot [color=black,solid,line width=1.0pt,mark=star,mark options={solid}]
  table[row sep=crcr]{%
1	4.34027777777778e-08\\
2	4.34027777777778e-08\\
4	8.98437500000027e-06\\
6	0.000160026041666667\\
8	0.000638498263888889\\
10	0.00153229166666667\\
};
\addlegendentry{$\text{SNR}_\text{OB}\text{=30dB}$};

\end{axis}
\end{tikzpicture}%
\end{subfigure}
\hspace{0.07\textwidth}
\begin{subfigure}[t]{0.23\textwidth}
\setlength{\tikzfigW}{0.9\textwidth}
\setlength{\tikzfigH}{0.9\textwidth}
\input{EveNDRnew.tikz}
\end{subfigure} \\
\begin{subfigure}[t]{0.22\textwidth}
\setlength{\tikzfigW}{0.8\linewidth}
\setlength{\tikzfigH}{0.8\linewidth}
\input{EveSNRnew.tikz}
\end{subfigure}
\hspace{0.02\textwidth}
\begin{subfigure}[t]{0.22\textwidth}
\setlength{\tikzfigW}{0.8\linewidth}
\setlength{\tikzfigH}{0.8\linewidth}
\input{attacknew.tikz}
\end{subfigure}
\hspace{0.01\textwidth}
\begin{subfigure}[t]{0.22\textwidth}
\setlength{\tikzfigW}{0.8\textwidth}
\setlength{\tikzfigH}{0.8\textwidth}
\input{AoD.tikz}
\end{subfigure}
\hspace{0.015\textwidth}
\begin{subfigure}[t]{0.22\textwidth}
\setlength{\tikzfigW}{0.8\textwidth}
\setlength{\tikzfigH}{0.8\textwidth}
% Minimal pgfplots version: 1.3
%  http://www.mathworks.com/matlabcentral/fileexchange/22022-matlab2tikz
\begin{tikzpicture}

\begin{axis}[%
width=\tikzfigW,
height=\tikzfigH,
scale only axis,
xmin=10,
xmax=40,
xtick={10, 20, 30, 40},
xlabel={No. of training modes},
ymode=log,
ymin=0.001,
ymax=1,
yminorticks=true,
ylabel={SER},
label style={font=\small},
tick label style={font=\small},
legend style={at={(0.51,0.65)},anchor=south west,legend cell align=left,align=left,draw=white!15!black,font=\scriptsize}
]
\addplot [color=black,solid,line width=1.0pt]
  table[row sep=crcr]{%
10	0.00756944444444445\\
20	0.00340277777777778\\
30	0.00277777777777778\\
40	0.00215277777777778\\
};
\addlegendentry{Bob};

\addplot [color=black,dashed,line width=1.0pt]
  table[row sep=crcr]{%
10	0.747309027777778\\
20	0.742795138888889\\
30	0.74296875\\
40	0.739756944444445\\
};
\addlegendentry{Eve};

\end{axis}
\end{tikzpicture}%
\end{subfigure} \\
\begin{subfigure}[t]{0.22\textwidth}
\includegraphics[width=\textwidth]{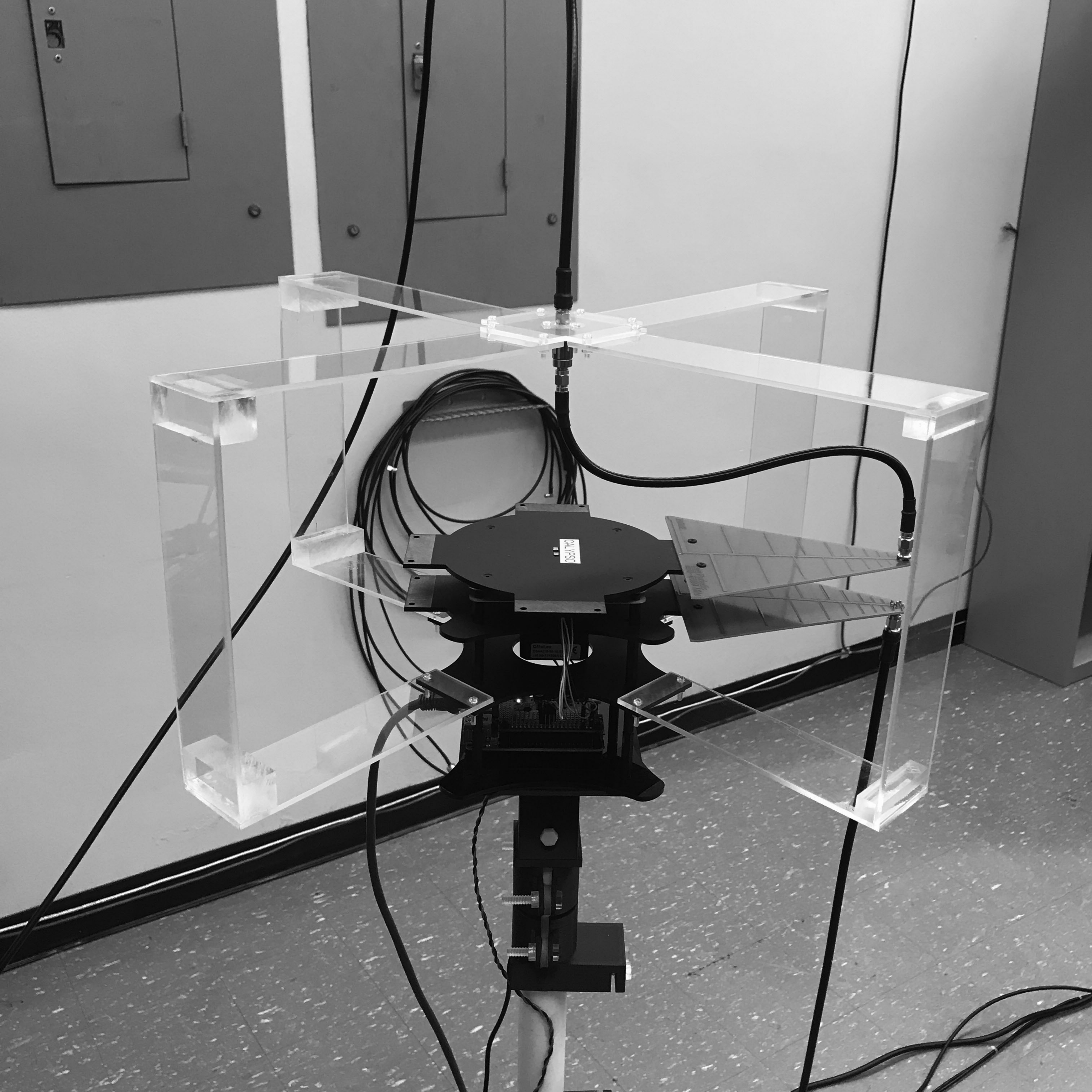}
\end{subfigure}
\hspace{\fill}
\begin{subfigure}[t]{0.22\textwidth}
\includegraphics[width=\textwidth]{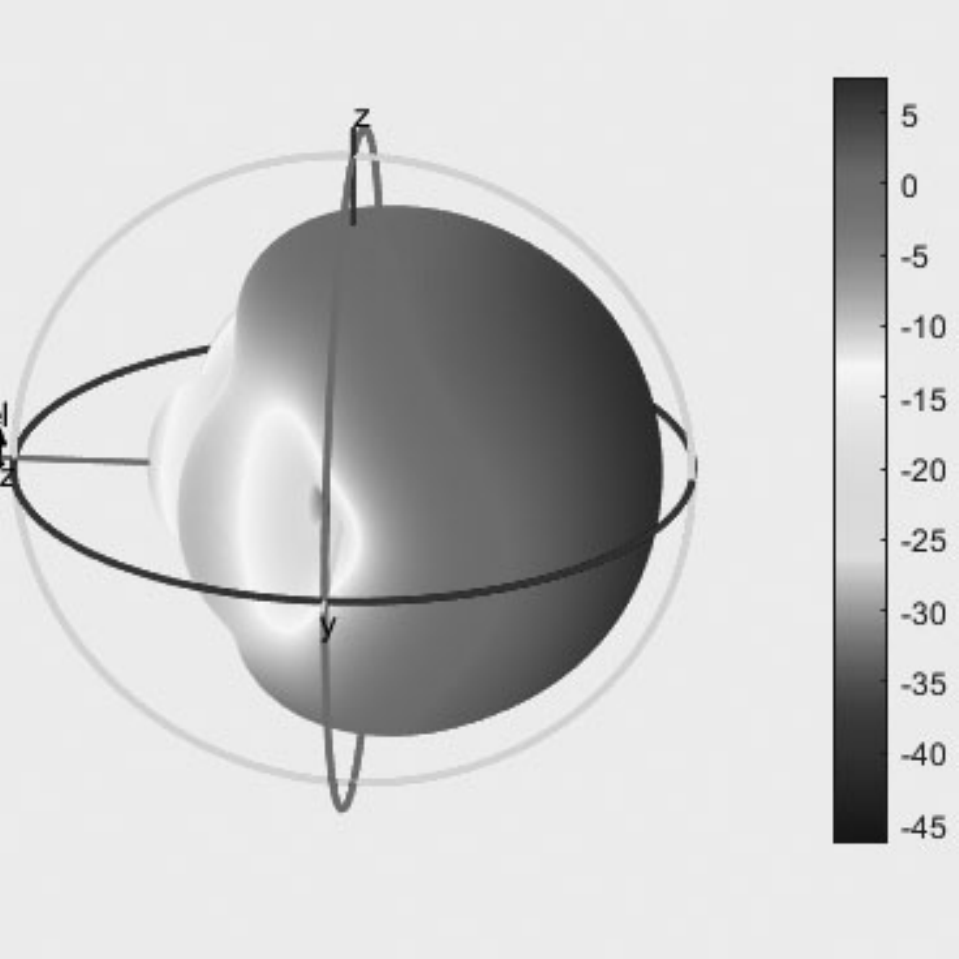}
\end{subfigure}
\hspace{\fill}
\begin{subfigure}[t]{0.22\textwidth}
\includegraphics[width=\textwidth]{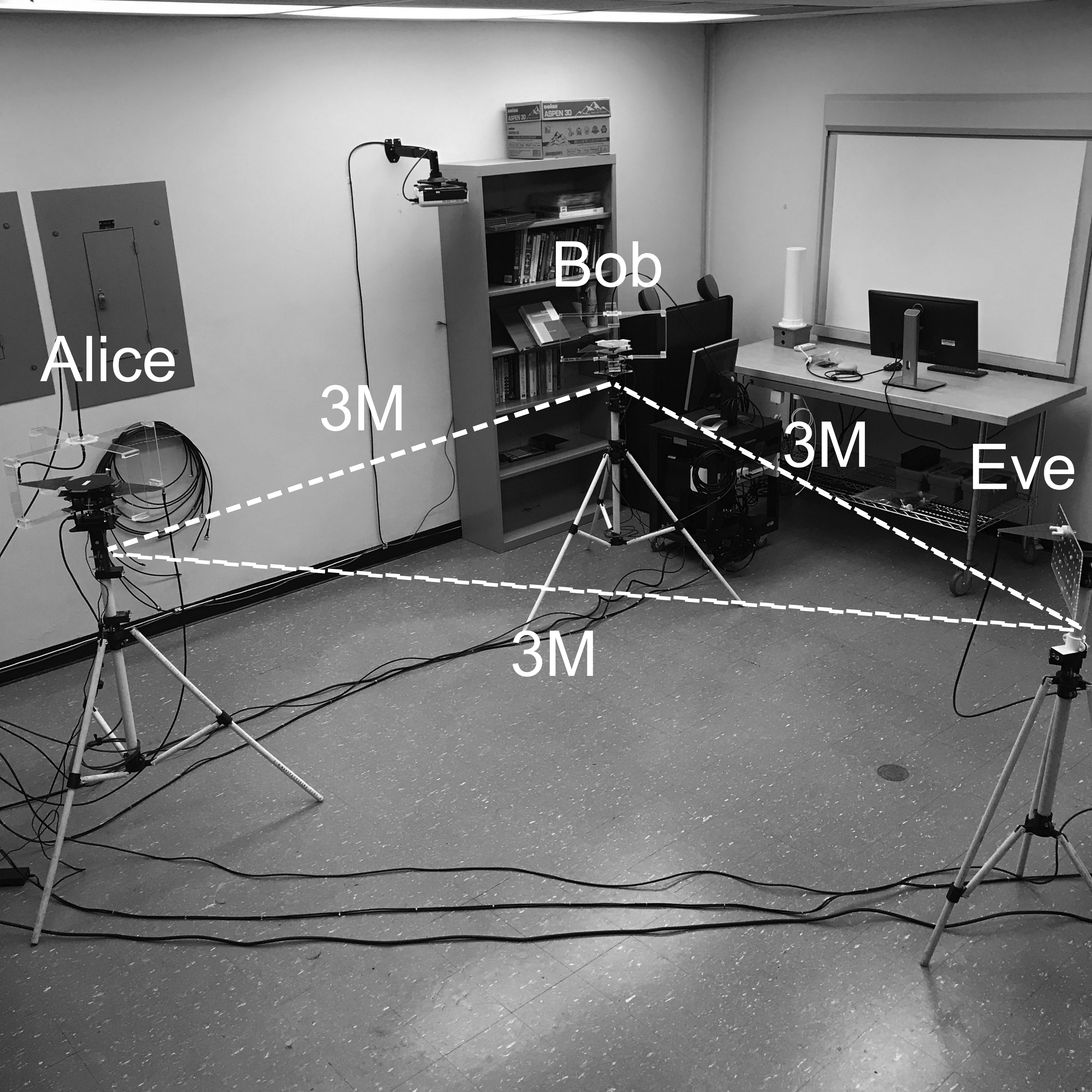}
\end{subfigure}
\hspace{\fill}
\begin{subfigure}[t]{0.22\textwidth}
\includegraphics[width=\textwidth]{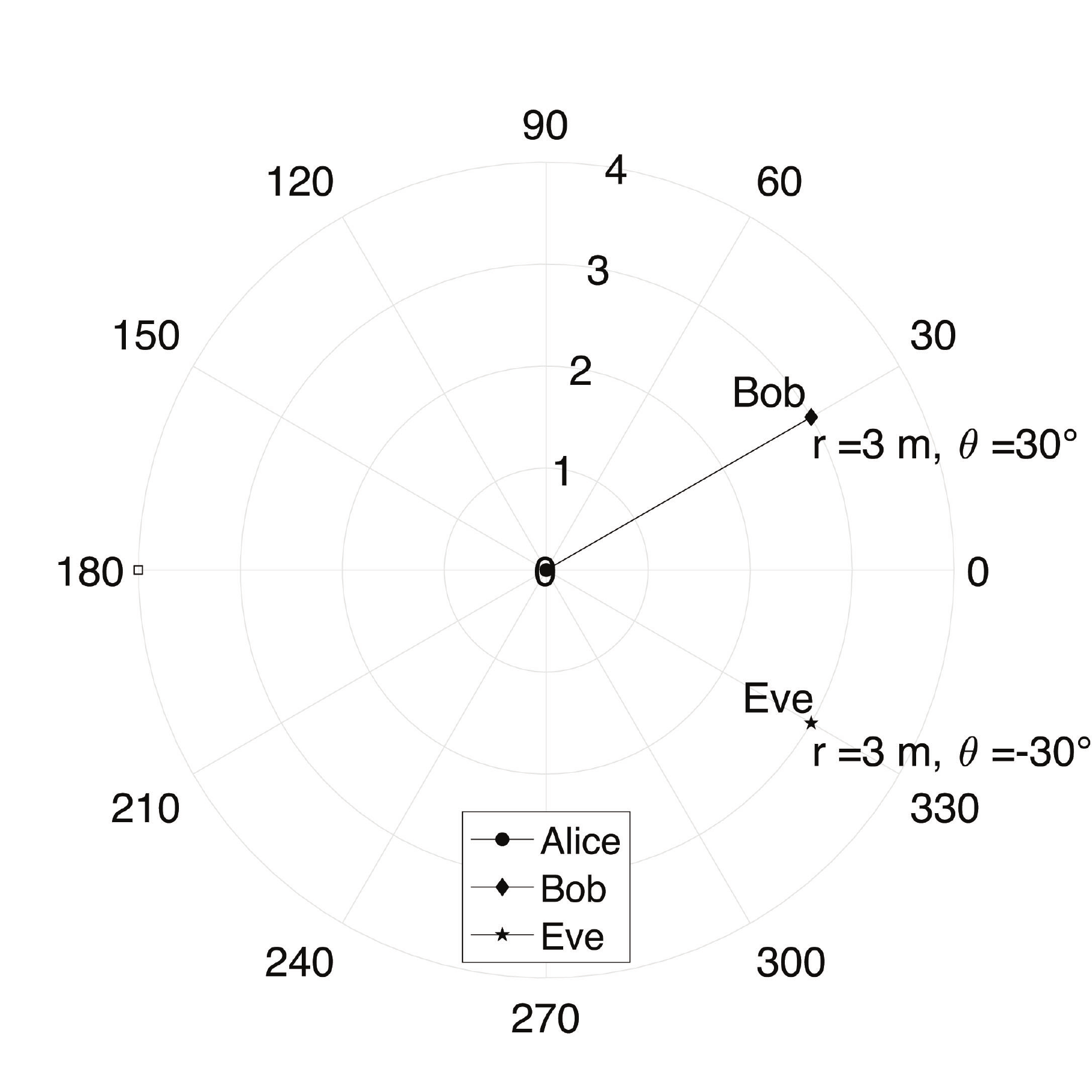}
\end{subfigure}
\caption{From top to bottom, left to right: (a) SER of Bob and Eve over the number of training modes. SNR = 25dB; NDR = 1. (b) SER of Bob over Alice's NDR for different SNRs, with 20 training modes. (c) Eve's SER over the number of iterations; SNR = 25dB; various NDR. (d) Eve's SER over the number of iterations. NDR = 1; various SNR. (e) Eve's SER over the number of iterations; SNR = 25dB; NDR = 1; various antenna switching period. (f) Azimuth CSI magnitude distribution estimated with compressive
sensing algorithm. (g) SER of Bob and Eve over the number of training mode based on the real-world channel data. (h) Real-world rotator. (i) Radiation pattern of the RA. (j) Real-world experiment setup. (k) Geometry of the setup.}
\label{fig:performance}
\end{figure*}

In this section, we evaluate the performance of our  scheme under the practical known-plaintext attack with both simulation and real-world experiments. We start with the overview of simulation setup, and investigate the effects of various parameters. With simulation, we can cover a wide parameter range and establish the operating environment for the known-plaintext attack with a MIMO eavesdropper. Then with an implementation using USRP platform and a rotating RA, we validate the simulation results via experiments. 

\subsection{Customized Reconfigurable Antenna}
For the channel randomization purpose, we prefer RAs with distinct radiation patterns across different antenna modes. There are different types of RAs in the literature \cite{bernhard2007reconfigurable,LiBeamSteeringReconfigurableAntenna2015}, however, most of them are designed for communication purpose that only steer to several directions, which results in similar radiation patterns over antenna modes and makes them unsuitable for channel randomization purpose. To better evaluate our ROBin scheme, we build our own RA by rotating a log periodic antenna manufactured by Ettus Research \cite{antenna}. We first measure all the design parameters for the given log periodic antenna, including arm width, arm spacing and etc., then its radiation pattern is simulated using MALTAB antenna toolbox and illustrated in Fig. \ref{fig:performance}i. In the simulation, we rotate the antenna every one degree, so that we have 360 antennas modes in total. And in practice, the rotator is constructed with a motor and  a microcontroller, to rotate the antenna agilely to an arbitrary angle in the azimuth plane. The rotator is illustrated in Fig. \ref{fig:performance}h. Note that, we can have various antenna configurations at Alice's side when Alice has multiple antennas in general, e.g. Alice can enrich antenna patterns by varying the gain level of each antenna, which can be achieved with power allocation among RF chains. Also, Alice can have more antennas than use and randomly selects one among them for transmission, or randomize the power ratio among antennas to introduce additional randomness to wireless channels as in \cite{hassanieh2015securing}.

% Hassanieh et al. in  \cite{hassanieh2015securing} also constructed a similar platform by rotating the transmitter with a fan motor at a constant speed. The transmitter is equipped with eight antennas, and each time a micro-controller randomly activates an antenna for transmission. However, for an eavesdropper who is able to measure the CSI of possible paths, the constant speed rotation allows him/her to predict the positions of antennas. In this way, the attacker can brute force the corresponding CSI of eight antennas to decode the message. In contrast, we rotate with a random speed \textcolor{red}{not sure}

\subsection{Simulation Setup}
As described in the system model, Alice, Bob and Eve are multi-antenna users with OFDM transmitters. W.l.o.g, we focus our simulation on a setup where Alice has two given log periodic antennas, Bob and Eve have one and two omnidirectional antenna(sTFor data transmission, the 30MHz wide AWGN channel is split into 48 equally spaced sub-channels, and the OFDM frames contain 192 symbols for each sub-channel. To evaluate the effect of Alice's AN, we vary the ratio of AN to the transmitted data signal, namely that Noise to Data Ratio (NDR). With fixed transmit power, the power for data signal is:
\begin{equation}
    D = \frac{1}{\text{NDR}+1}\begin{pmatrix}
        D_B\\
        \text{NDR} \cdot \text{AN}
    \end{pmatrix} 
\end{equation}
We simulate 100 different environment settings, where five scatters are put for each of them and the data signal are transmitted as 4-QAM symbols. The distance from Eve to Bob is set as $150$cm, which is 12 times of the signal wavelength. For all the simulations, we consider a more practical eavesdropper than that in theoretical analysis, where not all the historical data signal are known to the eavesdropper. Then we define the switching period $T$ of RA based on an OFDM frame, and 120 frames are sent during the channel coherent time, hence we have $T \in [1,120]$. For each frame, the attacker is assumed to obtain two symbols. 
Then if $T = 10$, it means that the transmission mode changes every ten packets, hence for a given transmit filter (computed from the given transmission mode), the attacker has 20 known symbols for filter training.
% where the symbols known by attackers are from the well-known protocols in the WiFi frame. Though no pilots are needed in transmission phase of ROBin, the attacker is still able to guess parts of symbols from information like addresses. However, the attacker is not able to obtain all the historical data signal in practice. Besides, since AN is not intended for the receiver, it is random data in general, which makes it hard for the attacker to get enough plaintext and identify the  correct  decoding  filter. Correspondingly, we define the switching period $T$ of RA based on an OFDM frame, and 120 frames are sent during the channel coherent time, hence we have $T \in [1,120]$. For each frame, the attacker is assumed to obtain two symbols. 
% Then if $T = 10$, it means that the transmission mode changes every ten packets, hence for a given transmit filter (computed from the given transmission mode), the attacker has 20 known symbols for filter training.

% \begin{figure}[t]
%     \Large
%     \centering
%     \resizebox{0.5\columnwidth}{!}{\input{numSamples.tikz}}
%     \caption{SER of Bob and Eve over the number of training modes. SNR = 25dB; NDR = 1; Eve's SER is obtained after 240 iterations.}
%     \label{fig:numSamples}
% \end{figure}
% \begin{figure}[t]
%     \Large
%     \centering
%     \resizebox{0.5\columnwidth}{!}{\input{SNR.tikz}}
%     \caption{SER of Bob over Alice's NDR for differnetn SNRs. Number of training modes is 20.}
%     \label{fig:SNRNDR}
% \end{figure}

\subsection{Effect of the number of training modes}
Since the estimation of the AoD distribution is based on compressive sensing in ROBin, theoretically, the more training modes we use, the more precise the estimation will be. Fig. \ref{fig:performance}a illustrates the SER of Bob and Eve over the number of training modes, obtained under ROBin and orthogonal blinding. Here to better show the impact of channel prediction to Bob's SER, we do not change the antenna mode during transmission, which is to set $T = 120$, then the only difference of these two schemes is that ROBin computes the transmit filter based on the predicted channel, while orthogonal blinding uses the measured channel matrix obtained from channel sounding. 
% With log scale, when SER is 0, we substitute it with the SER that only one symbol is incorrectly demodulated over all the experiments, which is $4.34 \times 10^{-8}$ in this case. 
From Fig. \ref{fig:performance}a we can see that there is a gap between Bob's SER obtained from two schemes, which is caused by the imperfect channel prediction and the missing channel sounding. However, it decreases with the number of training modes as expected, and when 20 training modes are used, Bob's SER is small enough for communication. On the other hand, Eve's SER obtained after 240 iterations is quite stable under the different number of training modes, this is because the effect of the transmit filter and artificial noise are both filtered out by Eve's receive filter.

\subsection{Effect of artificial noise and channel noise to Bob}
Fig. \ref{fig:performance}b shows Bob's SER with orthogonal blinding and ROBin, at this time the antenna switching period is set as $T = 6$, hence 20 transmission modes are used under a given environmental setting. From Fig. \ref{fig:performance}b, we can see that Bob's SER decreases as SNR increases under both schemes. Especially, both SNR and NDR have significant impacts on Bob's SER for orthogonal blinding. In contrast, the increase of SNR does not bring much benefit to Bob's SER for ROBin, since Bob's SER is dominated by the precise of channel prediction. Due to the imperfect channel prediction, part of the artificial noise is leaked to Bob's channel, which increases Bob's SER. Fortunately, as long as the NDR is not too large, the communication quality is still guaranteed. For instance, when SNR = 25dB and NDR = 2, Bob can still achieve an average SER of $1.1 \times 10^{-3}$. 

\subsection{Effect of artificial noise and channel noise to Eve}
Theoretically, the higher the NDR is, the higher is the SER on Eve's side. Here we set the antenna switching period as $T = 60$ to provide Eve some advantages. In Fig. \ref{fig:performance}c, we illustrate how Alice's NDR affects Eve's performance. As we expected, Eve's SER decreases with the increase of NDR. It is worth noticing that, when the power of artificial noise is not too strong, $\text{NDR} \leq 4$ for instance, we can see that Eve's SER has an obvious reduction with the iterative process; whereas, as the artificial noise becomes stronger, even if the number of iterations increases, the decrease of Eve's SER is not significant. 
Besides, in Fig. \ref{fig:performance}d we illustrate how SNR affects Eve's SER. The effect of channel noise to Eve's SER is much weaker than that to Bob's SER, no significant variation for Eve's SER with the increase of SNR. Hence we can conclude that Eve's attack performance is mainly constrained by the power of artificial noise that Alice sent. And there is a tradeoff between the system secrecy (Eve's SER) and the communication quality (Bob's SER) when injecting artificial noise to the channel.

% \begin{figure}[t]
%     \Large
%     \centering
%     \resizebox{0.5\columnwidth}{!}{\input{EveNDRnew.tikz}}
%     \caption{Eve's SER over the number of iterations; SNR = 25dB; various NDR.}
%     \label{fig:NDR_Eve}
% \end{figure}

% \begin{figure}[t]
%     \Large
%     \centering
%     \resizebox{0.5\columnwidth}{!}{\input{EveSNRnew.tikz}}
%     \caption{Eve's SER over the number of iterations. NDR = 1. various SNR.}
%     \label{fig:SNR_Eve}
% \end{figure}

\subsection{Effect of switching period to Eve}
Intuitively, the faster the antenna switches, the higher is Eve's SER. In Fig. \ref{fig:performance}e, we show Eve's SER over the antenna switching period. As we expect, Eve's SER decreases as $T$ increases. When $T = 60$, it is the best case for Eve under ROBin scheme in Fig. \ref{fig:performance}e, we can see that Eve's SER ($0.4047$) is still fairly high. To quantify ROBin's security improvement, we compute the difference between Eve's SER in ROBin and in orthogonal blinding and normalize it with Eve's SER in the worst case, e.g., the SER of random guessing. For instance, when Alice transmits QPSK (4QAM) symbols, we compute: $(\text{SER}_{\text{ROBin}}-\text{SER}_{\text{OB}})/0.75$. The result shows we can elevate the eavesdropper's SER by 46\% under 4-QAM modulation. When the antenna mode changes rapidly, especially for $T = 1$, we suppress Eve's attack success rate to the level of random guessing. Finally, we vary the number of known symbols in each frame from 2 to 20. And the result in Fig. \ref{fig:performance}e shows that Eve's SER does not fluctuate much due to the convergence of the algorithm.

% And to see how much performance we raise with ROBin in terms of Eve's SER, we normalize the increment of Eve's SER with her worst case, which is the random guessing with SER being 0.75, i.e., we compute: $(\text{SER}_{\text{E}}^{\text{OB+}} - \text{SER}_{\text{E}}^{\text{OB}}) / 0.75$, which turns out that, we can elevate the eavesdropper's SER by $46\%$. Besides, we can see that when the antenna mode changes rapidly, especially for $T = 1$, we suppress Eve's attack success rate to the level of random guessing. We also simulate how the number of known symbols in each frame affects Eve's performance, for which we increase it from 2 to 20. And the result shows that Eve's SER does not fluctuate much due to the convergence of the algorithm.

\begin{table}[t]
    \centering
    \caption{Secrecy leakage computed with real-world CSI}
    \begin{tabular}{|c|c|c|c|c|}
        \hline
        $|\mathcal{S}_1|$ & 10 & 20 & 30 & 40 \\
        \hline
        Secrecy Leakage & 0.36 & 0.23 & 0.22 & 0.21\\
        \hline
    \end{tabular}
    \label{tab:I}
\end{table}

\subsection{Effect of real-world channels}
We rotate the RA with the platform in Fig. \ref{fig:performance}h for the real-world CSI measurement in 2.6HGz. In the experiment, each of our OFDM frames contains 320 symbols and lasts for 0.08s. We collect the CSI data for about 60 seconds and change mode every $1/18$ seconds, which means the antenna switching period is less than the duration of a frame. To facilitate our simulation process, we set the antenna switching period to $T = 1$ while simulating. Based on the measuerd CSI, we first show the accuracy of the AoD estimation with 40 training modes in \ref{fig:performance}f. Observed that the predicted and measured CSI are similar for most of antennas modes, and our analysis result shows that the average prediction error decreases with the increase of training modes, which validates the effectiveness of our compressive sensing based AoD estimation algorithm. For all the parameters presented in simulations, we only show the SER of Bob and Eve over the number of training modes due to the page limitation. Observed that Fig. \ref{fig:performance}f shows a similar trend as in Fig. \ref{fig:performance}a, which indicates the consistency of our simulation and implementation. The secrecy we defined in Sec. \ref{sec:theory} is calculated with the measured CSI and shown in Table \ref{tab:I}, where $|\mathcal{S}_1|$ is the number of training modes. We can see that the secrecy leakage is low, however, it is nonzero though the SER of Eve is close to random guessing. This is because the secrecy leakage considers the temporal and spatial correlations of the main channel and the eavesdropper's channel, while the known-plaintext attack strategy only utilizes the temporal correlation of  the main channel.

%-------------------- Conclusion --------------------
\section{Conclusions}
In this paper, we propose an orthogonal blinding based secret transmission scheme, which is resistant to  known-plaintext attack by leveraging reconfigurable antennas  to rapidly randomize the channel state during transmission. We propose a compressive sensing based AoD estimation algorithm and  predict Alice-Bob  channel under  arbitrary antenna modes. We formally analyze the secrecy leakage using conditional mutual information, which is applicable to both single and multi-antenna eavesdroppers. We show that the secrecy leakage decreases with less channel correlation created by  artificial    channel randomization. We conduct extensive simulations and real-world experiments to evaluate its performance. Results show that, the communication quality between Alice and Bob remains acceptable, whereas the randomized channel can successfully prevent Eve from   launching the known-plaintext attack even if all the historical symbols are known. In the future, we will study other better practical alternatives to orthogonal blinding and analyze the security using secrecy capacity based notions.

%-------------------- Reference --------------------
\bibliographystyle{IEEEtran}
\bibliography{refpaper}

\begin{thebibliography}{10}
\providecommand{\url}[1]{#1}
\csname url@samestyle\endcsname
\providecommand{\newblock}{\relax}
\providecommand{\bibinfo}[2]{#2}
\providecommand{\BIBentrySTDinterwordspacing}{\spaceskip=0pt\relax}
\providecommand{\BIBentryALTinterwordstretchfactor}{4}
\providecommand{\BIBentryALTinterwordspacing}{\spaceskip=\fontdimen2\font plus
\BIBentryALTinterwordstretchfactor\fontdimen3\font minus
  \fontdimen4\font\relax}
\providecommand{\BIBforeignlanguage}[2]{{%
\expandafter\ifx\csname l@#1\endcsname\relax
\typeout{** WARNING: IEEEtran.bst: No hyphenation pattern has been}%
\typeout{** loaded for the language `#1'. Using the pattern for}%
\typeout{** the default language instead.}%
\else
\language=\csname l@#1\endcsname
\fi
#2}}
\providecommand{\BIBdecl}{\relax}
\BIBdecl

\bibitem{negi2005secret}
R.~Negi and S.~Goel, ``Secret communication using artificial noise,'' in
  \emph{IEEE vehicular technology conference}, vol.~62, no.~3.\hskip 1em plus
  0.5em minus 0.4em\relax Citeseer, 2005, p. 1906.

\bibitem{goel2008guaranteeing}
S.~Goel and R.~Negi, ``Guaranteeing secrecy using artificial noise,''
  \emph{IEEE transactions on wireless communications}, vol.~7, no.~6, pp.
  2180--2189, 2008.

\bibitem{liao2010qos}
W.-C. Liao, T.-H. Chang, W.-K. Ma, and C.-Y. Chi, ``Qos-based transmit
  beamforming in the presence of eavesdroppers: An optimized
  artificial-noise-aided approach,'' \emph{IEEE Transactions on Signal
  Processing}, vol.~59, no.~3, pp. 1202--1216, 2010.

\bibitem{li2011safe}
Q.~Li, W.-K. Ma, and A.~M.-C. So, ``Safe convex approximation to outage-based
  miso secrecy rate optimization under imperfect csi and with artificial
  noise,'' in \emph{2011 Conference Record of the Forty Fifth Asilomar
  Conference on Signals, Systems and Computers (ASILOMAR)}.\hskip 1em plus
  0.5em minus 0.4em\relax IEEE, 2011, pp. 207--211.

\bibitem{anand2012strobe}
N.~Anand, S.-J. Lee, and E.~W. Knightly, ``Strobe: Actively securing wireless
  communications using zero-forcing beamforming,'' in \emph{2012 Proceedings
  IEEE INFOCOM}.\hskip 1em plus 0.5em minus 0.4em\relax IEEE, 2012, pp.
  720--728.

\bibitem{argyraki2013creating}
K.~Argyraki, S.~Diggavi, M.~Duarte, C.~Fragouli, M.~Gatzianas, and
  P.~Kostopoulos, ``Creating secrets out of erasures,'' in \emph{Proceedings of
  the 19th annual international conference on Mobile computing \&
  networking}.\hskip 1em plus 0.5em minus 0.4em\relax ACM, 2013, pp. 429--440.

\bibitem{schulz2014practical}
M.~Schulz, A.~Loch, and M.~Hollick, ``Practical known-plaintext attacks against
  physical layer security in wireless mimo systems.'' in \emph{NDSS}, 2014.

\bibitem{ZhengHighlyEfficientKnownPlaintext2015}
Y.~Zheng, M.~Schulz, W.~Lou, Y.~T. Hou, and M.~Hollick, ``Highly efficient
  known-plaintext attacks against orthogonal blinding based physical layer
  security,'' \emph{IEEE Wireless Communications Letters}, vol.~4, no.~1, pp.
  34--37, 2014.

\bibitem{zheng2016profiling}
------, ``Profiling the strength of physical-layer security: A study in
  orthogonal blinding,'' in \emph{Proceedings of the 9th ACM Conference on
  Security \& Privacy in Wireless and Mobile Networks}.\hskip 1em plus 0.5em
  minus 0.4em\relax ACM, 2016, pp. 21--30.

\bibitem{wyner1975wire}
A.~D. Wyner, ``The wire-tap channel,'' \emph{Bell Labs Technical Journal},
  vol.~54, no.~8, pp. 1355--1387, 1975.

\bibitem{csiszar1978broadcast}
I.~Csisz{\'a}r and J.~Korner, ``Broadcast channels with confidential
  messages,'' \emph{IEEE transactions on information theory}, vol.~24, no.~3,
  pp. 339--348, 1978.

\bibitem{leung1978gaussian}
S.~Leung-Yan-Cheong and M.~Hellman, ``The gaussian wire-tap channel,''
  \emph{IEEE transactions on information theory}, vol.~24, no.~4, pp. 451--456,
  1978.

\bibitem{parada2005secrecy}
P.~Parada and R.~Blahut, ``Secrecy capacity of simo and slow fading channels,''
  in \emph{Information Theory, 2005. ISIT 2005. Proceedings. International
  Symposium on}.\hskip 1em plus 0.5em minus 0.4em\relax IEEE, 2005, pp.
  2152--2155.

\bibitem{li2007secret}
Z.~Li, W.~Trappe, and R.~Yates, ``Secret communication via multi-antenna
  transmission,'' in \emph{Information Sciences and Systems, 2007. CISS'07.
  41st Annual Conference on}.\hskip 1em plus 0.5em minus 0.4em\relax IEEE,
  2007, pp. 905--910.

\bibitem{gopala2008secrecy}
P.~K. Gopala, L.~Lai, and H.~El~Gamal, ``On the secrecy capacity of fading
  channels,'' \emph{IEEE Transactions on Information Theory}, vol.~54, no.~10,
  pp. 4687--4698, 2008.

\bibitem{KhistiSecureTransmissionMultiple2010}
A.~Khisti and G.~W. Wornell, ``Secure transmission with multiple antennas i:
  The misome wiretap channel,'' \emph{IEEE Transactions on Information Theory},
  vol.~56, no.~7, pp. 3088--3104, 2010.

\bibitem{KhistiSecureTransmissionMultiple2010a}
------, ``Secure transmission with multiple antennas—part ii: The mimome
  wiretap channel,'' \emph{IEEE Transactions on Information Theory}, vol.~11,
  no.~56, pp. 5515--5532, 2010.

\bibitem{gollakota2011they}
S.~Gollakota, H.~Hassanieh, B.~Ransford, D.~Katabi, and K.~Fu, ``They can hear
  your heartbeats: non-invasive security for implantable medical devices,'' in
  \emph{ACM SIGCOMM Computer Communication Review}, vol.~41, no.~4.\hskip 1em
  plus 0.5em minus 0.4em\relax ACM, 2011, pp. 2--13.

\bibitem{shen2013ally}
W.~Shen, P.~Ning, X.~He, and H.~Dai, ``Ally friendly jamming: How to jam your
  enemy and maintain your own wireless connectivity at the same time,'' in
  \emph{2013 IEEE Symposium on Security and Privacy}.\hskip 1em plus 0.5em
  minus 0.4em\relax IEEE, 2013, pp. 174--188.

\bibitem{tippenhauer2013limitations}
N.~O. Tippenhauer, L.~Malisa, A.~Ranganathan, and S.~Capkun, ``On limitations
  of friendly jamming for confidentiality,'' in \emph{Security and Privacy
  (SP), 2013 IEEE Symposium on}.\hskip 1em plus 0.5em minus 0.4em\relax IEEE,
  2013, pp. 160--173.

\bibitem{aono2005wireless}
T.~Aono, K.~Higuchi, T.~Ohira, B.~Komiyama, and H.~Sasaoka, ``Wireless secret
  key generation exploiting reactance-domain scalar response of multipath
  fading channels,'' \emph{IEEE Transactions on Antennas and Propagation},
  vol.~53, no.~11, pp. 3776--3784, 2005.

\bibitem{hassanieh2015securing}
H.~Hassanieh, J.~Wang, D.~Katabi, and T.~Kohno, ``Securing rfids by randomizing
  the modulation and channel,'' in \emph{12th USENIX Symposium on Networked
  Systems Design and Implementation (NSDI 15)}, 2015, pp. 235--249.

\bibitem{hou2015message}
Y.~Hou, M.~Li, R.~Chauhan, R.~M. Gerdes, and K.~Zeng, ``Message integrity
  protection over wireless channel by countering signal cancellation: Theory
  and practice,'' in \emph{Proceedings of the 10th ACM Symposium on
  Information, Computer and Communications Security}.\hskip 1em plus 0.5em
  minus 0.4em\relax ACM, 2015, pp. 261--272.

\bibitem{pan2017message}
Y.~Pan, Y.~Hou, M.~Li, R.~M. Gerdes, K.~Zeng, M.~A. Towfiq, and B.~A. Cetiner,
  ``Message integrity protection over wireless channel: countering signal
  cancellation via channel randomization,'' \emph{IEEE Transactions on
  Dependable and Secure Computing}, 2017.

\bibitem{bernhard2007reconfigurable}
J.~T. Bernhard, ``Reconfigurable antennas,'' \emph{Synthesis lectures on
  antennas}, vol.~2, no.~1, 2007.

\bibitem{rodrigo2014frequency}
D.~Rodrigo, B.~A. Cetiner \emph{et~al.}, ``Frequency, radiation pattern and
  polarization reconfigurable antenna using a parasitic pixel layer,''
  \emph{IEEE transactions on antennas and propagation}, vol.~62, no.~6, pp.
  3422--3427, 2014.

\bibitem{popper2011investigation}
C.~P{\"o}pper, N.~O. Tippenhauer, B.~Danev, and S.~Capkun, ``Investigation of
  signal and message manipulations on the wireless channel,'' in \emph{European
  Symposium on Research in Computer Security}.\hskip 1em plus 0.5em minus
  0.4em\relax Springer, 2011, pp. 40--59.

\bibitem{schmidt1986multiple}
R.~Schmidt, ``Multiple emitter location and signal parameter estimation,''
  \emph{IEEE transactions on antennas and propagation}, vol.~34, no.~3, pp.
  276--280, 1986.

\bibitem{ghassemzadeh2004measurement}
S.~S. Ghassemzadeh, R.~Jana, C.~W. Rice, W.~Turin, and V.~Tarokh, ``Measurement
  and modeling of an ultra-wide bandwidth indoor channel,'' \emph{IEEE
  Transactions on Communications}, vol.~52, no.~10, pp. 1786--1796, 2004.

\bibitem{czink2007cluster}
N.~Czink, X.~Yin, H.~Ozcelik, M.~Herdin, E.~Bonek, and B.~H. Fleury, ``Cluster
  characteristics in a mimo indoor propagation environment,'' \emph{IEEE
  Transactions on Wireless Communications}, vol.~6, no.~4, pp. 1465--1475,
  2007.

\bibitem{candes2008introduction}
E.~J. Cand{\`e}s and M.~B. Wakin, ``An introduction to compressive sampling [a
  sensing/sampling paradigm that goes against the common knowledge in data
  acquisition],'' \emph{IEEE signal processing magazine}, vol.~25, no.~2, pp.
  21--30, 2008.

\bibitem{xie2015hekaton}
X.~Xie, E.~Chai, X.~Zhang, K.~Sundaresan, A.~Khojastepour, and S.~Rangarajan,
  ``Hekaton: Efficient and practical large-scale mimo,'' in \emph{Proceedings
  of the 21st Annual International Conference on Mobile Computing and
  Networking}.\hskip 1em plus 0.5em minus 0.4em\relax ACM, 2015, pp. 304--316.

\bibitem{tan2000first}
C.~C. Tan and N.~C. Beaulieu, ``On first-order markov modeling for the rayleigh
  fading channel,'' \emph{IEEE Transactions on Communications}, vol.~48,
  no.~12, pp. 2032--2040, 2000.

\bibitem{LiBeamSteeringReconfigurableAntenna2015}
Z.~Li, E.~Ahmed, A.~M. Eltawil, and B.~A. Cetiner, ``A beam-steering
  reconfigurable antenna for wlan applications,'' \emph{IEEE Transactions on
  Antennas and Propagation}, vol.~63, no.~1, pp. 24--32, 2014.

\bibitem{antenna}
A.~Inc., ``Lp0965 antenna,''
  \url{https://www.ettus.com/product/details/LP0965}.

\end{thebibliography}

%-------------------- Appendix --------------------
\newpage
\section{Appendix}
\subsection{Proof of Theorem VI.1.}
\begin{proof}
To prove Theorem VI.1., which states
\begin{align*}
    I\left(\mathbf{D}_B(T);\mathbf{R}_E(T) \mid \right. & \left. \mathcal{H}_{AB}(T-1),\mathcal{H}_{AE}(T)\right) =  \nonumber \\
    I\left(\mathbf{D}_B(T);\mathbf{R}_E(T) \mid \right. & \left. \mathcal{H}_{ABE}\left(T-2\right), \delta \mathcal{H}_{ABE}(T) \right) = \nonumber \\
    I\left(\mathbf{D}_B(T);\mathbf{R}_E(T) \mid \right. & \left. \delta \mathcal{H}_{ABE}(T)\right),
\end{align*}
where 
\begin{align*}
\mathcal{H}_{ABE}(T-2) = \left\{ \right. & \left. \mathcal{H}_{AB}\left(T-2\right), \right. \left. \mathcal{H}_{AE}\left(T-2\right) \right. \}\\
\delta \mathcal{H}_{ABE}(T) = \left\{ \right. & \left. h_{AB}\left(T-1, u(T-1)\right), \right. \\
& \left. h_{AE}\left(T-1, u(T-1)\right), \right.\\
& \left. h_{AE}\left(T, u(T)\right)\ \right\}
\end{align*}
itis equivalent to prove that given $\delta \mathcal{H}_{ABE}(T)$, $\mathcal{H}_{ABE}(T-2)$ and $\left(\mathbf{D}_B(T),\mathbf{R}_E(T)\right)$ are conditionally independent. Since the messages $\mathbf{D}_B$ are independent from all the CSI information, and $\mathbf{R}_E(T)$ is a function of $\mathbf{D}_B(T)$ and $h_{AB}^{-1}\left(T,u(T)\right)$ plus some independent additive white Gaussian noise (AWGN), it is similar to prove that given $\delta \mathcal{H}_{ABE}(T)$, $\mathcal{H}_{ABE}(T-2)$ and $h_{AB}^{-1}(T,u(T))$ are conditionally independent, based on the Markov property.

Recall the Markov property of the channels:
\begin{equation*}
\begin{split}
& \Pr\left[h_{AB}\left(T, u(T)\right)\mid \mathcal{H}_{AB}(T-1)\right] = \\
& \ \ \ \ \Pr\left[h_{AB}\left(T, u(T) \right) \mid h_{AB}\left(T-1, u(T-1)\right)\right].
\end{split}
\end{equation*}
and
\begin{equation*}
\begin{split}
& \Pr\left[h_{AB}\left(T,u(T)\right)\mid \mathcal{H}_{AE}(T)\right] = \\
& \ \ \ \ \Pr\left[h_{AB}\left(T, u(T) \right) \mid h_{AE}\left(T, u(T)\right)\right]
\end{split}
\end{equation*}
To simplify, we denote $X_1 = \mathcal{H}_{AB}(T-2)$, $X_2 = h_{AB}(T-1,u(T-1))$, $X_3 = h_{AB}(T,u(T))$, and similarly define $Y$ for channel A-E. Then the Markov property can be rewritten as:
\begin{align*}
    \Pr(X_3|X_1,X_2) & = \Pr(X_3|X_2)\\
    \Pr(X_3|Y_1,Y_2,Y_3) & = \Pr(X_3|Y_3)
\end{align*}
which is illustrated below:
\begin{center}
\begin{tabular}{ c c c c c}
 $X_1$ & $\longrightarrow$ & $X_2$ & $\longrightarrow$ & $X_3$ \\ 
 $\big\updownarrow$ &  & $\big\updownarrow$ &  & $\big\updownarrow$ \\
 $Y_1$ & $\longrightarrow$ & $Y_2$ & $\longrightarrow$ & $Y_3$
\end{tabular}
\end{center}
And the CSI can be represented with $X$ and $Y$ in a simpler way as:
\begin{align*}
\mathcal{H}_{ABE}(T-2) & = \left\{ \right.  \left. X_1, \right. \left. Y_1 \right. \}\\
\delta \mathcal{H}_{ABE}(T) & = \left\{ \right.  \left. X_2, \right. \left. Y_2, \right. \left. Y_3 \right. \}\\
h_{AB}^{-1}\left( T,u(T)\right) & =   \left.X_3^{-1} \right.
\end{align*}
Hence our problem is equivalent to prove that given $(X_2,Y_2,Y_3)$, $(X_1,Y_1)$ and $X_3$ (which is equivalent to $X_3^{-1}$) are conditionally independent. Then we begin with
\begin{subequations}
    \begin{align*}
        &\Pr(X_1,Y_1,X_3|X_2,Y_2,Y_3)\\
        =&\Pr(X_3|X_2,Y_2,Y_3) \Pr(X_1,Y_1|X_2,Y_2,X_3,Y_3) \numberthis \label{eq9}
    \end{align*}
\end{subequations}
\eqref{eq9} is obtained by expressing the joint probability with the conditional probability, then we focus on simplifying its last term, for which we look at:
\begin{subequations}
    \begin{align*}
        &\Pr(X_3,Y_3|X_1,X_2,Y_1,Y_2)\\
        =&\Pr(X_3|X_1,X_2,Y_1,Y_2)\Pr(Y_3|X_1,X_2,Y_1,Y_2,X_3) \numberthis \label{eq1}\\
        =&\Pr(X_3|X_2)\Pr(Y_3|Y_2,X_3) \numberthis \label{eq2}\\
        =&\Pr(X_3|X_2,Y_2)\Pr(Y_3|X_2,Y_2,X_3) \numberthis \label{eq2_1}\\
        =&\Pr(X_3,Y_3|X_2,Y_2) \numberthis \label{eq2_2}\\
    \end{align*}
\end{subequations}
Similarly, \eqref{eq1} is obtained by expressing the joint probability with the conditional probability. With Markov property of the channels,
it is further simplified to \eqref{eq2}. Then we can add more conditional independent variables to it and get \eqref{eq2_1}, which equals to \eqref{eq2_2}. \eqref{eq2_2} implies that given $(X_2,Y_2)$, $(X_1,Y_1)$ and $(X_3,Y_3)$ are conditionally independent. Then back to \eqref{eq9}, we have 
\begin{subequations}
    \begin{align*}
        &\Pr(X_1,Y_1,X_3|X_2,Y_2,Y_3)\\
        =&\Pr(X_3|X_2,Y_2,Y_3) \Pr(X_1,Y_1|X_2,Y_2,X_3,Y_3) \numberthis \label{eq9_1}\\
        =&\Pr(X_3|X_2,Y_2,Y_3)\Pr(X_1,Y_1|X_2,Y_2) \numberthis \label{eq10}\\
        =&\Pr(X_3|X_2,Y_2,Y_3)\Pr(X_1,Y_1|X_2,Y_2,Y_3) \numberthis \label{eq11}
    \end{align*}
\end{subequations}
which means given $(X_2,Y_2,Y_3)$, $X_3$ and $(X_1,Y_1)$ are conditionally independent. Since $X_3^{-1}$ is a function of $X_3$, then this conditional independence still holds when we replace $X_3$ with $X_3^{-1}$, which implies:
\begin{equation}
    \Pr(X_3^{-1}|X_1,Y_1,X_2,Y_2,Y_3) = \Pr(X_3^{-1}|X_2,Y_2,Y_3) 
    \label{eq:XY}
\end{equation}
Then we reverse $X$, $Y$ in \eqref{eq:XY} back to the CSI, which gives us:
\begin{align}
& \Pr\left[ h_{AB}^{-1}\left(T,u(T)\right) | \mathcal{H}_{ABE}(T-2), \delta \mathcal{H}_{ABE}(T) \right] = \nonumber \\
& \Pr\left[ h_{AB}^{-1}\left(T,u(T)\right) | \mathcal{H}_{AB}(T-1), \mathcal{H}_{AE}(T) \right] = \nonumber \\
& \Pr\left[ h_{AB}^{-1}\left(T,u(T)\right) | \delta \mathcal{H}_{ABE}(T) \right] \label{eq133}
\end{align}

To compute $I\left(\mathbf{D}_B(T);\mathbf{R}_E(T) \mid \right. \left. \mathcal{H}_{AB}(T-1),\mathcal{H}_{AE}(T)\right)$, we ignore the AWGN and consider $\Pr\left(\mathbf{D}_B(T);\hat{\mathbf{R}}_E(T) \mid \mathcal{H}_{AB}(T-1),\mathcal{H}_{AE}(T)\right)$
first, where $\hat{\mathbf{R}}_E(T) = h_{AB}^{-1}(\left(T,u(T)\right)  \mathbf{D}_B(T)$.

\begin{subequations}
    \begin{align*}
        &\Pr\left(\mathbf{D}_B(T),\hat{\mathbf{R}}_E(T) \mid \mathcal{H}_{AB}(T-1),\mathcal{H}_{AE}(T)\right)\\
        = &\Pr\left(\mathbf{D}_B(T),h_{AB}^{-1}\left(T,u(T)\right)=\frac{\hat{\mathbf{R}}_E(T)}{\mathbf{D}_B(T)}\mid \mathcal{H}_{AB}(T-1), \mathcal{H}_{AE}(T)\right) \numberthis \label{eq13}\\
        = &\Pr\left(\mathbf{D}_B(T)\mid \mathcal{H}_{AB}(T-1), \mathcal{H}_{AE}(T)\right) \\
        &\times \Pr\left(h_{AB}^{-1}\left(T,u(T)\right)=\frac{\hat{\mathbf{R}}_E(T)}{\mathbf{D}_B(T)}\mid \mathcal{H}_{AB}(T-1), \mathcal{H}_{AE}(T)\right) \numberthis \label{eq14} \\
        = &\Pr\left(\mathbf{D}_B(T)\mid \mathcal{H}_{ABE}(T)\right) \\
        & \ \ \times \Pr\left(h_{AB}^{-1}\left(T,u(T)\right)=\frac{\hat{\mathbf{R}}_E(T)}{\mathbf{D}_B(T)}\mid \delta\mathcal{H}_{ABE}(T)\right) \numberthis \label{eq15} \\
        = & \Pr\left(\mathbf{D}_B(T),\hat{\mathbf{R}}_E(T) \mid \delta\mathcal{H}_{ABE}(T)\right) \numberthis \label{eq16}
    \end{align*}
\end{subequations}
With the fact that messages are independent from all the CSI information, we can get \eqref{eq14}, and meanwhile get rid of $\mathcal{H}_{ABE}(T-2)$ from $\mathbf{D}_B$'s condition, which gives us the first term of \eqref{eq15}, and with \eqref{eq133} we get the second term of \eqref{eq15}. Then by converting conditional probability to joint probability, we get \eqref{eq16}. Since the AWGN is independent from every term of above equations, we can add it into $\hat{\mathbf{R}}_E$ and get $\mathbf{R}_E$ while above results still holds. 

So far, we have proved that 
\begin{align*}
    I\left(\mathbf{D}_B(T);\mathbf{R}_E(T) \mid \right. & \left. \mathcal{H}_{AB}(T-1),\mathcal{H}_{AE}(T)\right) =  \nonumber \\
    I\left(\mathbf{D}_B(T);\mathbf{R}_E(T) \mid \right. & \left. \delta \mathcal{H}_{ABE}(T)\right)
\end{align*}
for single antenna system. Next, we present the approach to extend it to MIMO. Note that, for the MIMO system, each element in the Markov chain becomes the channel matrix. Similarly,
\begin{subequations}
    \begin{align*}
        &\Pr\left(\mathbf{D}_B(T),\hat{\mathbf{R}}_E(T) \mid \mathcal{H}_{AB}(T-1),\mathcal{H}_{AE}(T)\right)\\
        = &\Pr\left(\mathbf{D}_B(T),\mathbf{H}_{AB}^{-1}\left(T,u(T)\right)\in\mathbf{\Gamma}\mid \mathcal{H}_{AB}(T-1), \mathcal{H}_{AE}(T)\right) \numberthis \label{eq13}
    \end{align*}
\end{subequations}
where
\begin{equation*}
    \mathbf{\Gamma} =\{\mathbf{H}_{AB}^{-1}\left(T,u(T)\right)\in\mathbf{\Gamma},\text{s.t.~} \hat{\mathbf{R}}_E(T) = \mathbf{H}_{AB}^{-1}(T,u(T))\mathbf{D}_B(T)\}
\end{equation*}
and represents a set of matrices where its element is a possible solution for $\mathbf{H}_{AB}^{-1}\left(T,u(T)\right)$. Then we can eliminate $\mathcal{H}_{ABE}(T-2)$ with similar procedures from Eq. \eqref{eq14} to Eq. \eqref{eq16}.  

\end{proof}

% \input{Intro}
% \input{Relatedwork}
% \input{Model}
% \input{Physicallayersecurity}
% \input{Attack}
% \input{Scheme}
% \input{Theory}
% \input{Performance}
% \input{Conclusion}
% % \input{Appendix}

% % \printbibliography
% \bibliographystyle{IEEEtran}
% % \bibliography{refpaper,references_y_zheng}
% \bibliography{refpaper}
% \input{Appendix3}

\end{document}